\documentclass[apj]{emulateapj}

\shorttitle{The BSS population in Cetus and Tucana}
\shortauthors{Monelli et al.}

\begin{document}

\title{The ACS LCID project VII: the blue stragglers population in the
isolated dSph galaxies Cetus and Tucana\altaffilmark{1}}


\author{M. Monelli\altaffilmark{2,3},
    S. Cassisi\altaffilmark{4},
    M. Mapelli\altaffilmark{5},
    E.J. Bernard\altaffilmark{6,2},
    A. Aparicio\altaffilmark{2,3},
    E.D. Skillman\altaffilmark{7},
    P.B. Stetson\altaffilmark{8},
    C. Gallart\altaffilmark{2,3},
    S.L. Hidalgo\altaffilmark{2,3},
    L. Mayer\altaffilmark{9,10}, and
    E. Tolstoy\altaffilmark{11}
    }
    
\altaffiltext{1}{Based on observations made with the NASA/ESA {\it Hubble Space
   Telescope}, obtained at the Space Telescope Science Institute, which is
    operated by the Association of Universities for Research in Astronomy,
    Inc., under NASA contract NAS5-26555. These observations are associated
    with program 10505.}
\altaffiltext{2}{Instituto de Astrof\'{i}sica de Canarias, La Laguna, Tenerife,
    Spain; monelli@iac.es, carme@iac.es, antapaj@iac.es, shidalgo@iac.es.}
\altaffiltext{3}{Departamento de Astrof\'{i}sica, Universidad de La Laguna, 
    Tenerife, Spain}
\altaffiltext{4}{INAF-Osservatorio Astronomico di Collurania,
    Teramo, Italy; cassisi@oa-teramo.inaf.it.}
\altaffiltext{5}{Universit\`a di Milano Bicocca, Dipartimento di Fisica 
G.Occhialini, Piazza della Scienza 3, I--20126, Milano, Italy;  mapelli@mib.infn.it }
\altaffiltext{6}{Institute for Astronomy, University of Edinburgh, Royal 
    Observatory, Blackford Hill, Edinburgh EH9 3HJ, UK; ejb@roe.ac.uk}
\altaffiltext{7}{Department of Astronomy, University of Minnesota,
    Minneapolis, USA; skillman@astro.umn.edu.}
\altaffiltext{8}{Dominion Astrophysical Observatory, Herzberg Institute of
    Astrophysics, National Research Council, Victoria, Canada;
    peter.stetson@nrc-cnrc.gc.ca.}
\altaffiltext{9}{Department of Physics, Institut f\"ur Astronomie,
    ETH Z\"urich, Z\"urich, Switzerland; lucio@phys.ethz.ch.}
\altaffiltext{10}{Institut f\"ur Theoretische Physik, University of Zurich,
    Z\"urich, Switzerland; lucio@physik.unizh.ch.}
\altaffiltext{11}{Kapteyn Astronomical Institute, University of Groningen,
    Groningen, Netherlands; etolstoy@astro.rug.nl.}

\begin{abstract}
We present the first investigation of the Blue Straggler star (BSS) population in two 
isolated dwarf spheroidal galaxies of the Local Group, Cetus and Tucana. Deep HST/ACS photometry
allowed us to identify samples of 940 and 1214 candidates, respectively. The analysis
of the star formation histories of the two galaxies suggests that both host
a population of BSSs.  Specifically, if the BSS candidates are interpreted as young 
main sequence stars, they do not conform to their galaxy's age-metallicity relationship. 
The analysis of the luminosity function and the radial distributions support this conclusion, 
and suggest a non-collisional mechanism for the BSS formation, from the evolution of
primordial binaries. This scenario is also supported by the results of new dynamical 
simulations presented here.  Both galaxies coincide with the relationship between the 
BSS frequency and the absolute visual magnitude M$_V$ found by \citet{momany07}.  
If this relationship is confirmed by larger sample, then it could be a valuable tool 
to discriminate between the presence of BSSs and galaxies hosting truly young populations.
\end{abstract}


\section{Introduction}\label{sec:intro}

Blue Straggler stars (BSSs) were first identified in the Galactic globular cluster (GGC)
M3 by \citet{sandage53}, as a group of stars bluer and brighter than the 
cluster turn-off stars, thus being more massive than the stars currently evolving at the
turn-off (TO). Since then, BSSs have been identified in a variety
of stellar systems in very different environmental conditions including: the field
\citep{carney05}, open clusters \citep{liu08}, globular clusters \citep[e.g.,][]{piotto04},
and dwarf galaxies \citep{lee03, momany07}. 
Their puzzling position in the color-magnitude diagram (CMD) suggested
that they do not fit in the traditional scheme of stellar evolution for single stars.
In fact, in order to populate that region of the CMD, these exotic objects must have
experienced a physical process able to allow them to still stay in the core H-burning 
stage despite of their mass and the cluster age \citep[for a detailed historical review, 
see][]{stryker93}.

Presently, two different physical mechanisms are favored for explaining the formation 
of BSSs: {\em i)} coalescence through direct stellar collisions \citep[COL-BS,][]{hills76}, 
and {\em ii)} evolution of primordial binary systems, in which mass 
transfer between the two components allows the rejuvenation of the secondary component 
\citep[MT-BSS,][]{mccrea64}. It is also widely accepted that these two scenarios 
are not in competition, but they could both occur 
simultaneously in a stellar system. In fact, the coexistence of BSSs 
formed through both the collisional and mass transfer channels is commonly invoked 
to interpret the radial distribution of BSSs in many GGCs. The higher 
central concentration is explained by an efficient rate of stellar collisions in 
the densest regions, while the rise at large distance would be due to MT-BSSs 
\citep{mapelli06, ferraro09} surviving in the outskirts. 

While the scenario for BSSs in Galactic clusters is well settled, 
at least from the observational point of view, 
in contrast, this is not true for dwarf galaxies. In fact, even though the first
indication for the presence of BSS candidates in the Sextans dwarf dates back to
\citet{mateo91}, and, since then, several observational findings suggest the
presence of BSS candidates in several dSphs \citep[see, e.g.,][and references
therein]{mapelli09}, we are still faced with the problem of understanding if 
these candidates are {\sl genuine} BSSs or {\sl normal} core H-burning stars 
belonging to an, albeit sparse, intermediate-age stellar population.

Recently, a few studies have focused on the properties of BSSs in nearby
dSph galaxies in the Local Group (LG). \citet{mapelli07} presented wide 
field data for the galaxies Draco and Ursa Minor, concluding that the population
of BSSs is, in both systems, compatible with the MT-BSS scenario. \citet{mapelli09} 
discussed the properties of BSSs in  Sculptor and Fornax. Sculptor presents 
similarities with both Draco and Ursa Minor, as its BSS population is also compatible
with a MT-BSS population. On the contrary, the relatively strong central radial 
concentration of the BSS candidate stars in Fornax favors the presence of an 
intermediate-age stellar population (in agreement with previous studies 
of the star formation history (SFH) of Fornax).

Unraveling the nature of the BSS candidates has important implications for
our understanding of the SFHs of these galaxies.
For example, 
whenever it is possible to discriminate between a sequence of BSSs and a young MS 
population, it is possible to set a constraint on the age of the last star formation
event in a galaxy. Therefore, the BSS populations in Draco, Ursa Minor, and Sculptor 
suggest that no star formation occurred in the last few Gyr in these galaxies. 

Intriguingly, \citet{momany07} found a statistically significant
anti-correlation between the frequency of BSSs and the absolute visual magnitude (M$_V$)
of 8 nearby dSphs. The anti-correlation is valid over a large range of luminosities, and
presents a different slope than that of GGCs. In addition, \citet{momany07} 
ruled out the possibility that collisional binaries could have contributed to the 
observed BSS populations and concluded that the BSSs in dwarf spheroidal galaxies 
are mainly formed via mass exchange in primordial binaries.

In view of the important implications of the presence of a sizable sample of BSSs 
in a dwarf galaxy, it is worthwhile to apply the approach outlined by \citet{momany07}
to other dwarfs.  The hope is to eventually produce independent constraints useful 
for addressing the true nature of the BSS candidates.



\begin{deluxetable}{l|cc}
\tabletypesize{\scriptsize}
\tablewidth{0pt}
\tablecaption{The distance modulus, reddening, and candidate BSS properties. \label{tab:tab01}}
\tablehead{
\colhead{---} & \colhead{Cetus} & \colhead{Tucana} }
\startdata
$(m-M)_0$          &   $24.46\pm0.12$  &    $24.74\pm0.12$	\\
$E(B-V)$           &      $0.028$      &         $0.031$	\\
N${_{BS}}^a$       &       $940$       &         $1214$	        \\
N${_{BS_{evol}}}^b$ &       $90$       &          $123$	        \\
F${_{BS}}^c$       &   $0.05\pm0.05$   &    $0.02\pm0.04$	\\
N$_{AC}^d$         &      $3$          &          $6$           \\   
\enddata
\tablenotetext{a}{Number of candidate BSSs}
\tablenotetext{b}{Number of evolved BSSs in the core He-burning stage}
\tablenotetext{c}{~Logarithmic frequency of BSSs with respect HB of stars}
\tablenotetext{d}{Number of Anomalous Cepheids}
\end{deluxetable}


In this paper, we investigate these issues further by analyzing the properties of 
candidate BSSs in the isolated dSphs Cetus and Tucana. The data sets are part of
the LCID project ({\it Local Cosmology from Isolated Dwarfs}), aimed at recovering
the full SFHs of six isolated LG galaxies: Cetus and Tucana (dSph), LGS~3 and Phoenix 
(dIrr/dSph), IC~1613 and Leo~A (dIrr). The plan of the paper is as follows: in the 
next section we briefly describe the photometric data set and outline the criteria for 
selecting the candidate BSSs in the observed CMDs. 
In \S 3 we investigate the properties of the BSS populations. 
\S 4 presents the results of dynamical simulations. 
The discussion and final remarks close the paper.

\section{The observations and the selection of samples}\label{sec:data}

The observations used in this paper have already been presented in \citet[][Cetus]{monelli10b}
and \citet[][Tucana]{monelli10c}. Here we summarize the main points relevant to the
following discussion. The images were collected with the ACS camera aboard the HST, under 
the project {\itshape The onset of star formation in the universe: constraints from nearby 
isolated dwarf galaxies} (PID 10505, PI C. Gallart). A total of 25 and 32 orbits have
been devoted to Cetus and Tucana, respectively, using the $F475W$ 
and $F814W$ filter pass bands. The photometric reduction has been performed using the 
DAOPHOT/ALLFRAME package \citep{alf}. The final catalog has been calibrated to
the VEGAMAG system using the transformations presented in \citet{sirianni05}.
The adopted distance modulus and reddening are summarized in Table \ref{tab:tab01}, 
together with other quantities calculated in this work.


\begin{figure}
\centering
 \includegraphics[width=8.5cm]{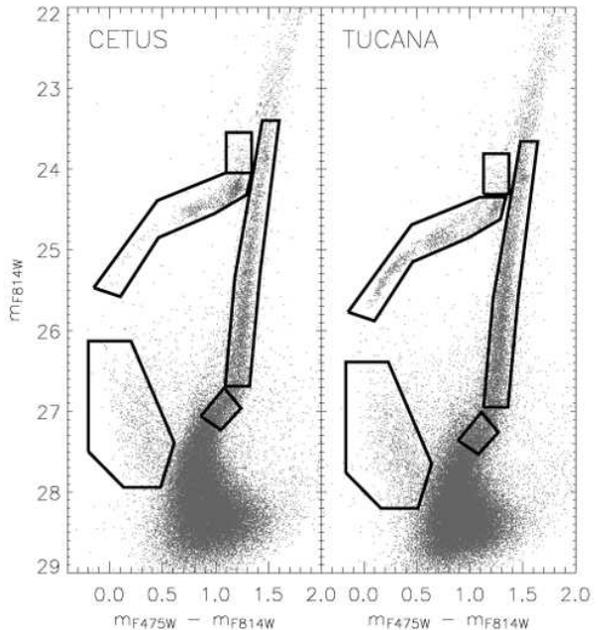}
 \caption{The observed CMDs for the two dwarfs: Cetus (left panel) and Tucana (right panel). 
 The delimited areas represent the regions of the CMDs selected for the present analysis 
 (see the text for more details).}
 \label{fig:samples}
\end{figure}


Figure \ref{fig:samples} shows the CMDs of the two galaxies, with the regions adopted 
to select the various samples studied superimposed. Note that we adopted the same boxes
for both galaxies, applying a shift to take into account the differences in distance modulus
and reddening. The CMDs of both galaxies are $\approx 1.5$ mag deeper than the TO, reaching 
$m_{F814W} \sim 28.8$ mag. The blue plume of candidate BSSs emerges clearly in the CMD, 
for $m_{F475W} - m_{F814W} < 0.5$ mag, $26 \la m_{F814W} \la 28$. We defined a box
in this region to include the bulk of these stars. Note that there are a number of 
objects between the BSSs region and the red giant branch (RGB). We decided not
to include these stars for two reasons: {\em i)} it is possible that blended
objects from the most populated TO region are polluting this part of the CMD;
{\em ii)} in this color range, we expect to find stars evolved off the main sequence.
Also, we did not include the brightest objects ($m_{F814W} \sim 26.2$) because it is
possible that they are extreme horizontal branch (HB) stars. Also note that we estimated 
a negligible contamination from both foreground Galactic stars (due both to the small 
area covered and the high Galactic latitude of the two galaxies) and background 
galaxies (thanks to the careful cleaning of the catalogs, see \citealt{monelli10b}).
Figure \ref{fig:samples} also shows the regions adopted to select stars in the HB, 
the RGB, and the progeny of BSSs during the central helium burning phase. This last
box has been placed above the red HB, paying attention not to include the AGB
clump, which clearly appears in the CMDs of both galaxies ($m_{F814W}
\sim 23.4$ mag).


\begin{figure*}
 \includegraphics[width=12cm]{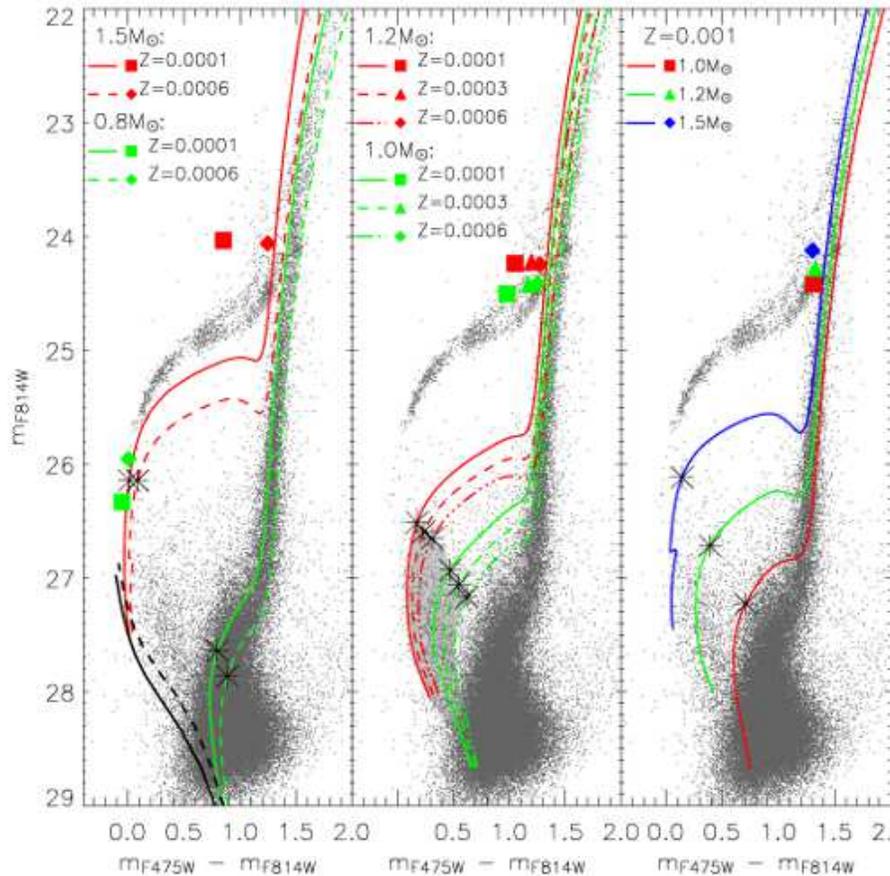}
 \vskip 0.8cm
 \caption{The CMD of Tucana with superimposed a selected sample of evolutionary 
 stellar models for the labeled assumptions about the initial stellar mass and 
 chemical compositions.  Note that, to avoid confusion, the tracks are plotted 
 until the tip of the RGB, and that the filled symbols mark the Helium ignition.
 We have adopted the values of distance modulus and reddening,
 from \citet{bernard08}, as summarized in Table \ref{tab:tab01}.
 {\itshape Left -} The 1.5 and 0.8{}$M_{\odot}$ bracket the
 entire sequence of candidate BSSs. The two black lines show the ZAMS for
 metallicity z=0.0001 (solid) and Z=0.001 (dashed), which outline the blue
 edge of the sequence.
  {\itshape Central -} We show the ranking  with metallicity of stars of 1.0 (green) 
 and 1.2 (red) {}$M_{\odot}$.  The shaded area encloses the central hydrogen 
 burning phase, until the exhaustion marked by the asterisks. 
 {\itshape Right -} Comparison with models with fixed metallicity (Z=0.001)
 and three different masses (1.0, 1.2, 1.5 {}$M_{\odot}$).}
 \label{fig:track}
\end{figure*}


\section{Analysis of the candidate BSS}\label{sec:analysis}

In the box covering the blue plume, we have identified 940 and 1214 objects
in Cetus and Tucana, respectively, which we consider as BSS candidates.
Note that the present data cover a significant fraction of the body of Tucana (tidal
radius $r_t = 3.45\arcmin$), which was centered in the ACS field (field-of-view $3.4\arcmin$).
However, Cetus occupies a larger area in the sky  ($r_t \sim 32\arcmin$, 
\citealt{mcconnachie06}\footnotemark[12]), and the sampled field covers a smaller fraction. 
\footnotetext[12]{New preliminary estimates based on wide-field Subaru data confirm
values $r_t > 15\arcmin$ (Bernard et al, 2012, in prep.), therefore significantly
larger than the ACS field-of-view.} Moreover, 
the observed field was offset from the center so that the innermost 
regions of Cetus were not observed \citep{monelli10b}. If we take this into account,
following the reasoning of \citet{bernard09} to estimate the total number 
of RR Lyrae stars,
the total number of BSS candidates in Cetus is estimated to be $\simeq$ 5,000 stars.

The most important question is to assess the nature of these stars. BSSs have been
found in very different environments, from the field, to low-density stellar systems 
such as open clusters and dwarf spheroidals, and in most globular clusters. They are 
therefore common products, and it is reasonable to expect to find them in Tucana and 
Cetus as well. Therefore, the basic question we are trying to address is: are
these candidates genuine BSSs, or is there also a significant component of truly 
young stars? In the following sections we will try to address this using different 
observables.


\begin{figure}
\hskip 0.4cm
 \includegraphics[width=8.5cm]{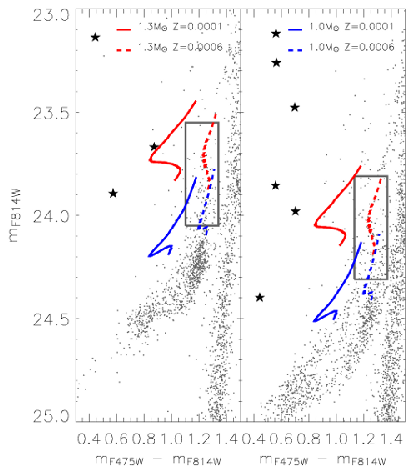}
 \caption{Similar to Figure.~\ref{fig:track}, but we present a zoom of the CMD of Cetus
 (left panel) and Tucana (right panel), corresponding to the core He-burning stage. In 
 each panel the box represents the region used to select the candidate progeny of the
 blue plume stars. Four tracks are overplotted, for different mass and metallicity assumptions.
 The filled stars mark the location of the anomalous Cepheid variables detected in these 
 galaxies \citep{bernard09}.
 }
  \label{fig:trackevol}
\end{figure}


	\subsection{Clues from the detailed SFH analysis}\label{sec:sfh}

 The main objective of the LCID project was to derive accurate SFHs for a sample
of six isolated galaxies in the Local Group, in order to understand their evolution
in a cosmological context. To do this, we adopted well-established techniques based
on the comparison of the observed CMD with a synthetic one. We devoted significant effort
to investigating the error budget and also possible systematics affecting our solution,
due to the use of different photometric packages, stellar evolutions libraries, SFH
codes. The main conclusion \citep[see][]{monelli10b,monelli10c,hidalgo11} of this
analysis was that we can rely, within the error bars, on both the uniqueness and stability
of the derived solutions. Therefore, the most important features recovered, such as
the epoch and the duration of the main star formation events, including those at the
oldest epoch, and the age-metallicity relations are solid results.

The SFHs presented in \citet{monelli10b} and \citet{monelli10c}, for Cetus and
Tucana respectively, show that both galaxies are made of old and 
metal-poor populations. In particular, both galaxies formed 90\% of their stars 
at epochs older than 9~Gyr ago, and the bulk of the stars have metallicities in the 
range $0.0001 < Z < 0.001$. In addition, both galaxies present a well defined 
age-metallicity relation in the sense of increasing metallicity with decreasing
age.  

The SFH reconstructions indicate a small population ($< 3\%$ in mass) 
of stars significantly younger than the majority of stars ($2 < t < 5$ Gyr)
but with metallicities consistent with the oldest stars ($Z < 0.0006$). 
Thus, these stars do not follow the general age-metallicity relations in these 
two galaxies. The analysis of the best-fit CMDs show that they nicely populate
the blue plume region of the CMD, brighter and bluer than the oldest MS turn-off.
The possibility that the metallicity of these stars has been understimated by the
code is excluded with high confidence level. In fact, higher metallicity 
and an age of $\approx 4$ Gyr, would produce also a redder RGB sequence that is 
not observed. On the basis of the retrieved SFH, \citet{monelli10b} made the hypothesis
that the sequence of blue stars, brighter than the old MS Turn-Off, are consistent
with a population of BSSs, characterized by the same metal content as
the metal-poor dominant stellar component.

An alternative interpretation would be that these are truly young metal-poor MS
stars, formed as a consequence of gas accretion. However, it seems very unlikely that 
two galaxies, that spent most of their life in isolation, accreted gas and formed stars
of the same metallicity and at the same epoch, in one event that created roughly the
same amount of mass (3\% of the total stellar mass).

	\subsection{Comparison with stellar evolutionary models}\label{sec:track}

Figure~\ref{fig:track} shows the comparison of the Tucana CMD with theoretical tracks
from the BaSTI database\footnotemark[13]\footnotetext[13]{\itshape http://www.oa-teramo.inaf.it/BASTI}
of stellar evolution models \citep{pietrinferni04}. Here we note that the
analysis of the SFH for the LCID galaxies has been performed by using the same
evolutionary framework; so the present analysis is fully consistent with those concerning the
SFHs. The selected tracks span the range of metallicities suggested by the SFHs, Z =
0.0001, 0.0003,  0.0006, and 0.001 as an upper limit. 

The figure discloses some interesting clues. First, the black lines in the left panel show
the Zero Age Main Sequence (ZAMS) for two metallicities: Z=0.0001 (solid line) and Z = 0.001
(dashed). The comparison with the observed CMDs shows that both ZAMSs border the
blue envelope of the BSS sequence, independent of the assumed metal content. 
Moreover, the same panel shows that the maximum mass fitting the brightest 
objects is equal to about 1.5 {}$M_{\odot}$ (red lines), again with a negligible 
dependence on the assumed metallicity (Z=0.0001, 0.001 for the red solid and dashed lines,
respectively). The same figure also shows the location of two theoretical tracks 
corresponding to $0.8M_\odot$ stellar models (green lines), taken as representative of 
the stellar structures currently evolving at the TO and sub-giant branch of both galaxies.
The filled symbols show the first point corresponding to the central helium burning
phase.
It is evident - as also fully supported by the SFH
analysis - that for the TO and the SGB, a range of metallicities in the adopted
evolutionary framework is required to properly interpret the observed morphology of the
CMD. This comparison reveals that the maximum mass  of the stars in the blue plume
is at most $\sim$ 2 times larger than the mass of the stars at the TO. This occurrence is 
in good agreement with the  properties of BSSs in other stellar 
systems such as GGCs \citep[see, i.e.,][and references therein]{gilliland98}.


\begin{figure*}
\centering
\includegraphics[width=8cm]{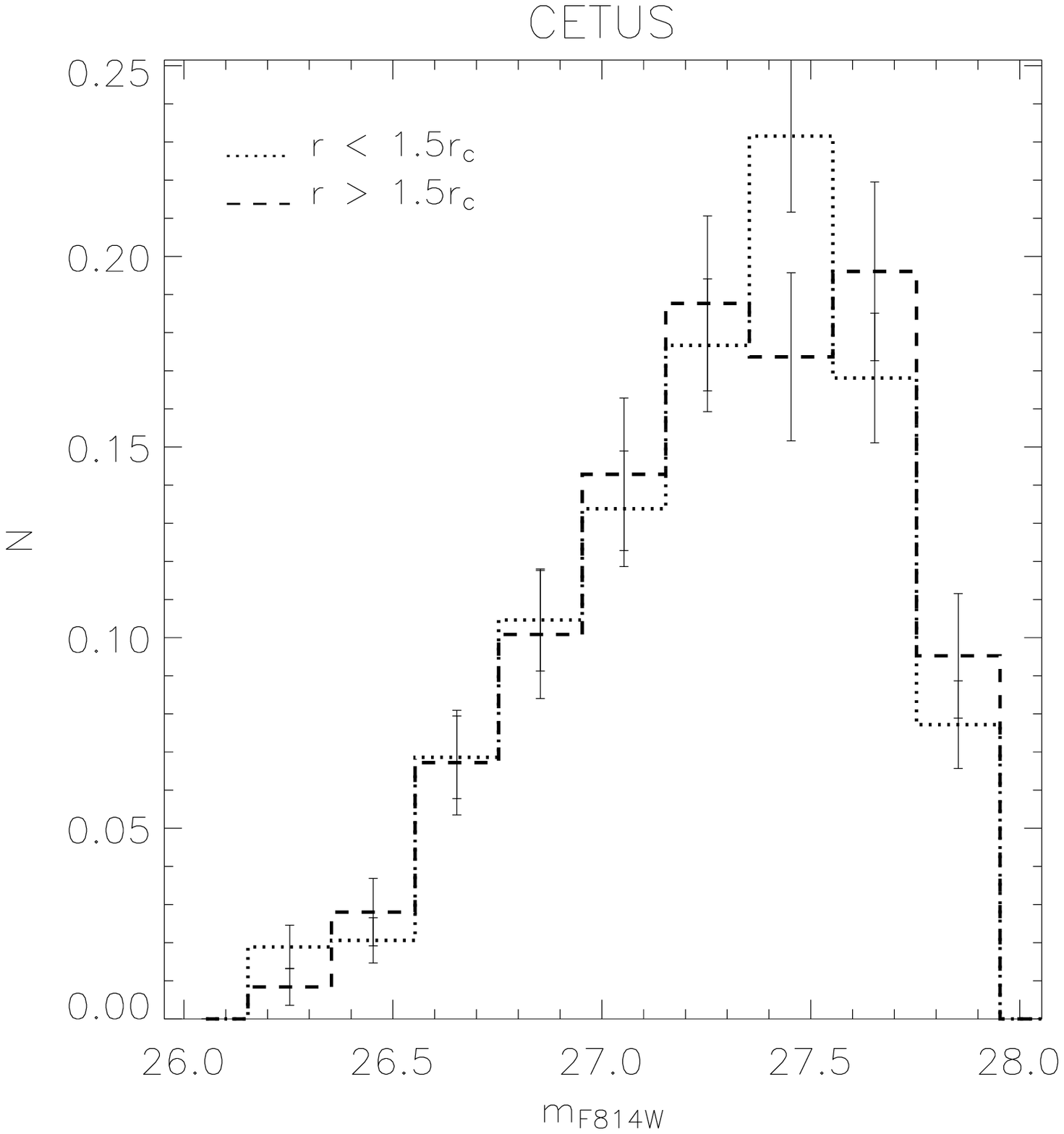}
\includegraphics[width=8cm]{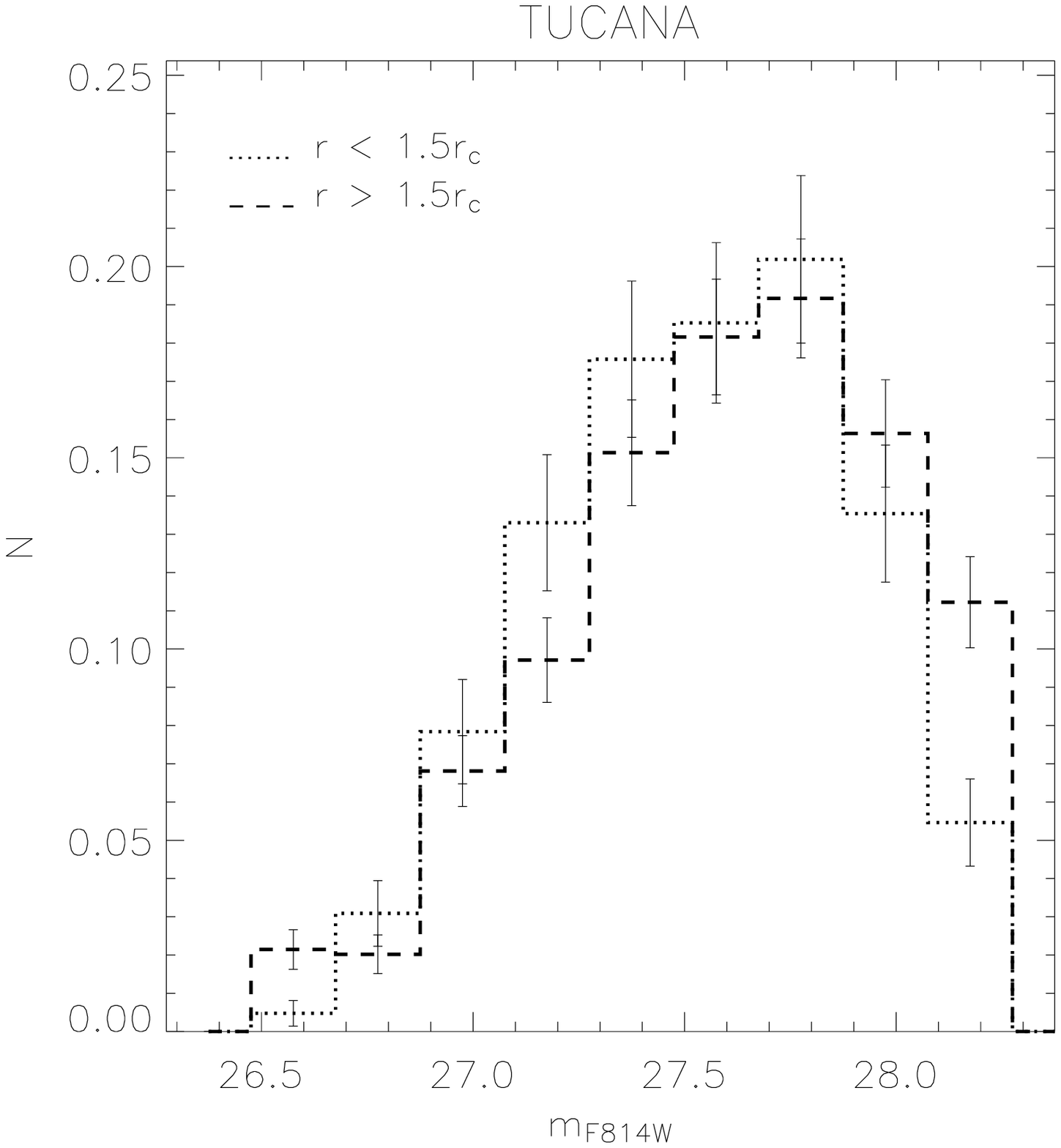}
\caption{Normalized luminosity function of the candidate BSS population in Cetus (left panel) 
and Tucana (right panel). The different lines refer to the BSS stars within (dotted) 
and outside (dashed) 1.5$r_c$ from the center.}
\label{fig:lumfun_rad}
\end{figure*}


The central panel shows the comparison with stellar tracks for different masses ($M = 1.0,
1.2 M_\odot$, green and red lines, respectively), and different metallicities 
 ($Z=0.0001$-solid, $Z=0.0003$-dashed, $Z=0.0006$-dot-dashed). The shaded area 
encloses the region corresponding to the central hydrogen burning phase, whose end is 
marked by the asterisks. The bulk of BSS candidates clearly occupies this region of the 
CMD.\footnotemark[14]\footnotetext[14]{Although the adopted stellar models are obtained in the canonical 
evolutionary framework, i.e., they do not account for the physical processes at the basis 
of BSS formation, we think that in order to have a rough estimate
of the metallicity range spanned by the BSS candidates, they are still suitable.}
The plot clearly shows that a spread in mass is needed to fit the observed distribution 
of  blue plume stars. A spread in metallicity is also possible but, as shown in the
right panel of the same figure, values larger than $Z > 0.001$ seem unlikely. The 1.0 $M_\odot$
track is too faint and too red to represent the bulk of the blue plume objects. The
1.2 $M_\odot$ still gives a good representation, and can be considered as an upper limit
both in term in of mass and metallicity. In fact, despite the 1.5 $M_\odot$ still
limits the brightest blue plume stars, the age of this model at the turn-off ($\sim$1.3Gyr)
is in contrast with the finding of the SFH. On the other hand, higher metallicities
would shift the tracks to redder color and fainter magnitude, significantly worsening
the agreement in particular in the RGB phase.
Thus, relatively metal-poor and relatively massive ($>1\ M_{\odot}$) stellar structures 
are preferred to better represent the peak of the distribution. 
Therefore, the metallicity derived for the BSS candidates is in good agreement with 
that estimated for the bulk of the 
older populations, as discussed in \S \ref{sec:sfh}, and not in agreement with the 
higher metallicities found at the end of the initial episodes of star formation in
Cetus and Tucana. This result can be considered a circumstantial evidence 
supporting the idea that - at least a large fraction of - the stars located along the 
blue plume could be genuine BSSs; in fact if they were true young stars formed in 
a late star formation burst, they would have a metallicity lower than that one could 
expect of the basis of a standard age-metallicity relation.

Additional information can be provided by the study of the BSS progeny in the core He-burning 
stage. Evolutionary models for MS stars predict that stars more massive than about $1M_{\odot}$ are
located in the red side of the HB at a magnitude level depending on their He core mass
\citep{castellani95}. In particular, the right-hand panel of Figure~\ref{fig:track} suggests 
that the evolved stars in this mass range all clump in the same region of the CMD.
This implies that the progeny of BSSs of similar mass, independent of the formation mechanism
\citep[see, e.g.,][]{sills09}, should occupy a similar region in the CMD, redder and brighter 
than their low-mass counterparts. This means that, on the one hand, the distribution of BSS progeny in
the CMD provides very little information about the mass distribution of the BSS population
above $1M_{\odot}$. On the other hand, the ratio of the number of stars can be compared with
theoretical expectations based on the evolutionary lifetimes in the core H- and He-burning stages.

Figure~\ref{fig:trackevol} shows an expanded view of the HB region of the Cetus (left panel) 
and Tucana (right panel) CMDs. Superimposed to both plots are the evolutionary tracks,
starting with the onset of the central helium burning,
for two masses (1.0 -blue line- and 1.3 -red-$M_{\odot}$) and two metallicities ($Z=0.0001$
-solid- and $Z=0.0006$ -dashed-). The boxes corresponding to our selections are shown.
Both panels show the presence of a sparse sequence above the red HB. While a few of 
them, which are located close to theoretical prediction for the most metal-poor stars, fall 
outside our box, the majority of stars, inside the box, look more compatible with
slightly more metal-rich tracks. Note that this is in agreement with the SFH results,
which gives for the two galaxies a mean metallicity of the order of $Z=0.0004$ (Cetus)
and $Z=0.0006$ (Tucana).

To further investigate the interpretation that the stars located inside the box showed in 
Fig.~\ref{fig:trackevol} can be really considered the progeny of the stars located 
along the MS blue plume, we compared the star counts ratio between the objects 
located inside the two boxes with theoretical predictions obtained
taking into the account the information from the SFH. \citet[][their Fig 18]{monelli10b}
have already shown a synthetic CMD built from the SFH solution of Cetus, where stars
with colors and magnitudes typical of a population of BSSs clearly appear.
We therefore counted the stars using the same boxes defined here, and estimated the 
same ratio using a best-fit CMD for both Cetus and Tucana. The derived ratios are 
0.08$\pm$0.01 and 0.08$\pm$0.01 for Cetus and Tucana respectively.
Analogous estimates can be made from the observed CMD. The number of BSSs in Cetus and 
Tucana are 940 and 1214, respectively, while the number of objects in the evolved 
phase are 90 and 123. This gives ratios of 0.10$\pm$0.01 for Cetus and 0.10$\pm$0.01 
for Tucana. This shows that there is good agreement with the empirical values.
This suggests that the selected stars are the evolved counterpart of blue plume objects.
Moreover, this test strongly suggests that, if we are dealing with a pure BSSs 
population, the analysis performed using models calculated for normal MS stars is only 
marginally affecting the results.


\begin{figure*}
\resizebox{8truecm}{7truecm}{\includegraphics{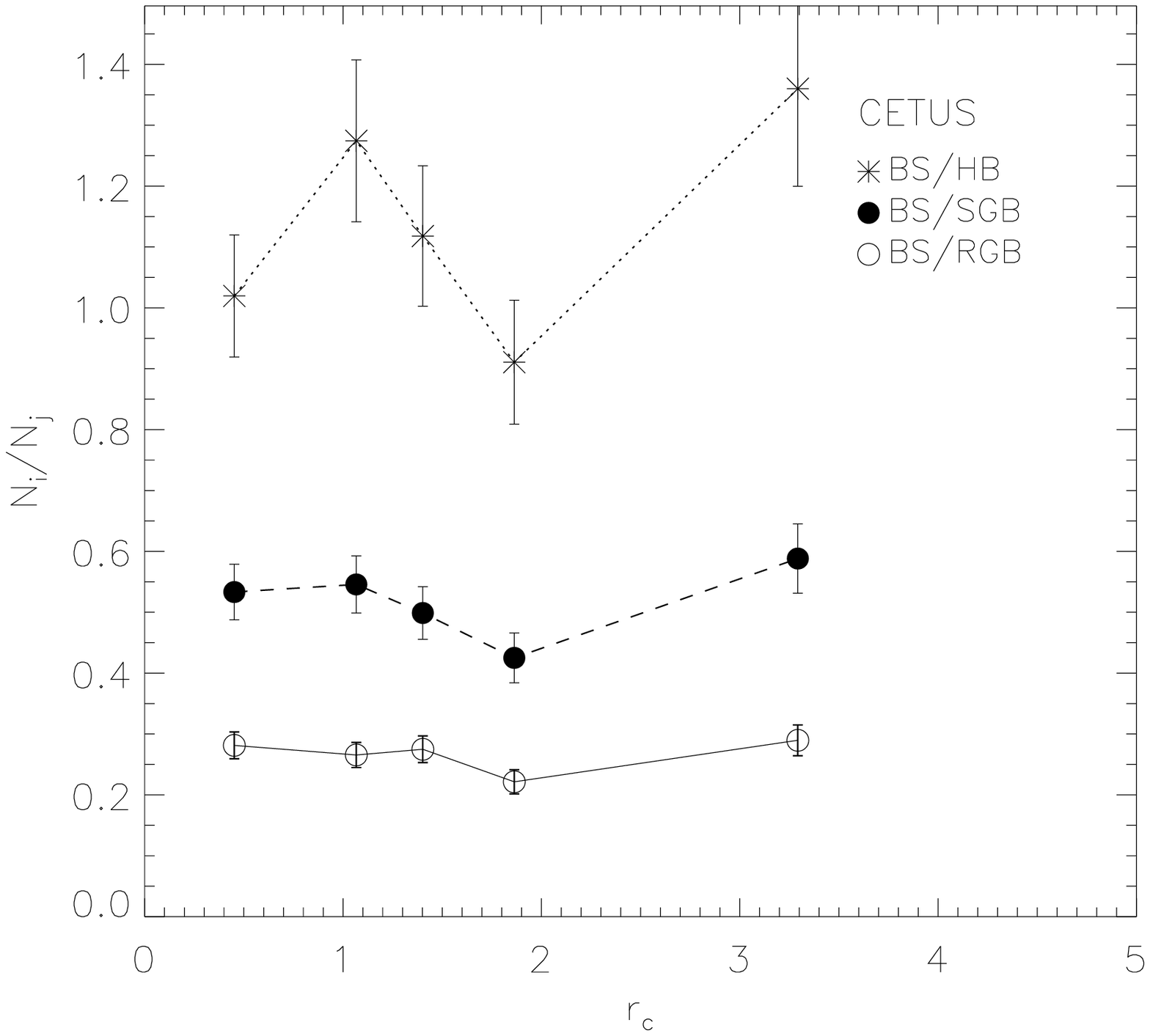}}
\resizebox{8truecm}{7truecm}{\includegraphics{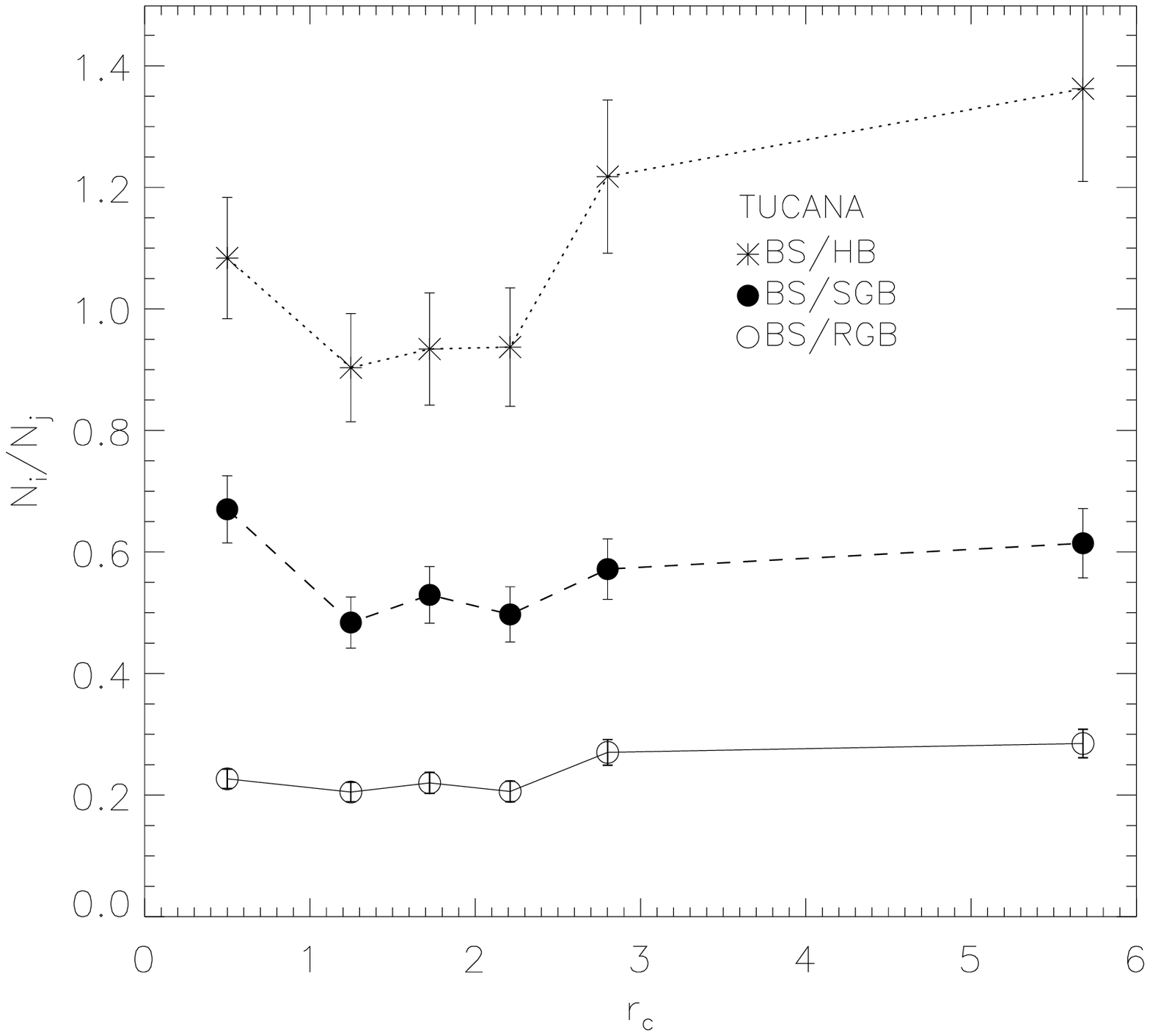}}
 \caption{Ratios of the number of BSSs with respect of HB (asterisks), SGB (full circles), 
 and RGB (open circles) stars, as a function of the galactocentric radius for Cetus
 (left) and Tucana (right).}
 \label{fig:grad_rad}
\end{figure*}


It is also worth noting that the progeny of BSS - in particular the more massive ones -
can cross the instability strip for radial pulsation during the core He-burning stage; in
this case they pulsate as anomalous Cepheids \citep{bono97}. When studying the
populations of variable stars in the LCID galaxies, \citet{bernard09} discovered three 
anomalous Cepheids in Cetus, and six in Tucana. The locations of these variable stars in 
the CMDs of both galaxies are shown in Fig. \ref{fig:trackevol}. They appear bluer and
brighter than the expected bulk of the BSS progeny. This suggests that anomalous Cepheids
might have slightly larger masses than the typical BSS.


\begin{figure*}
\resizebox{8truecm}{7truecm}{\includegraphics{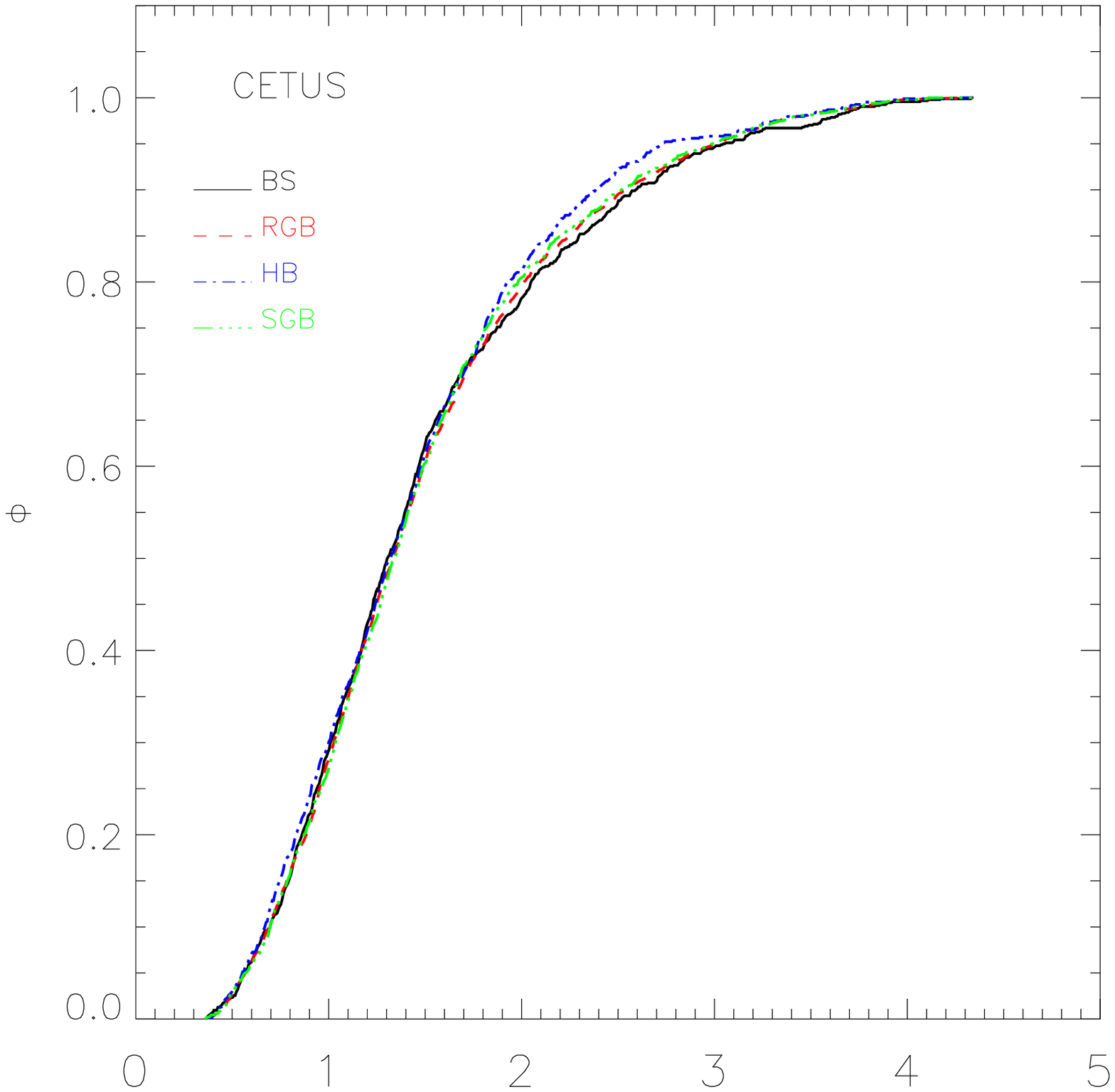}}
\resizebox{8truecm}{7truecm}{\includegraphics{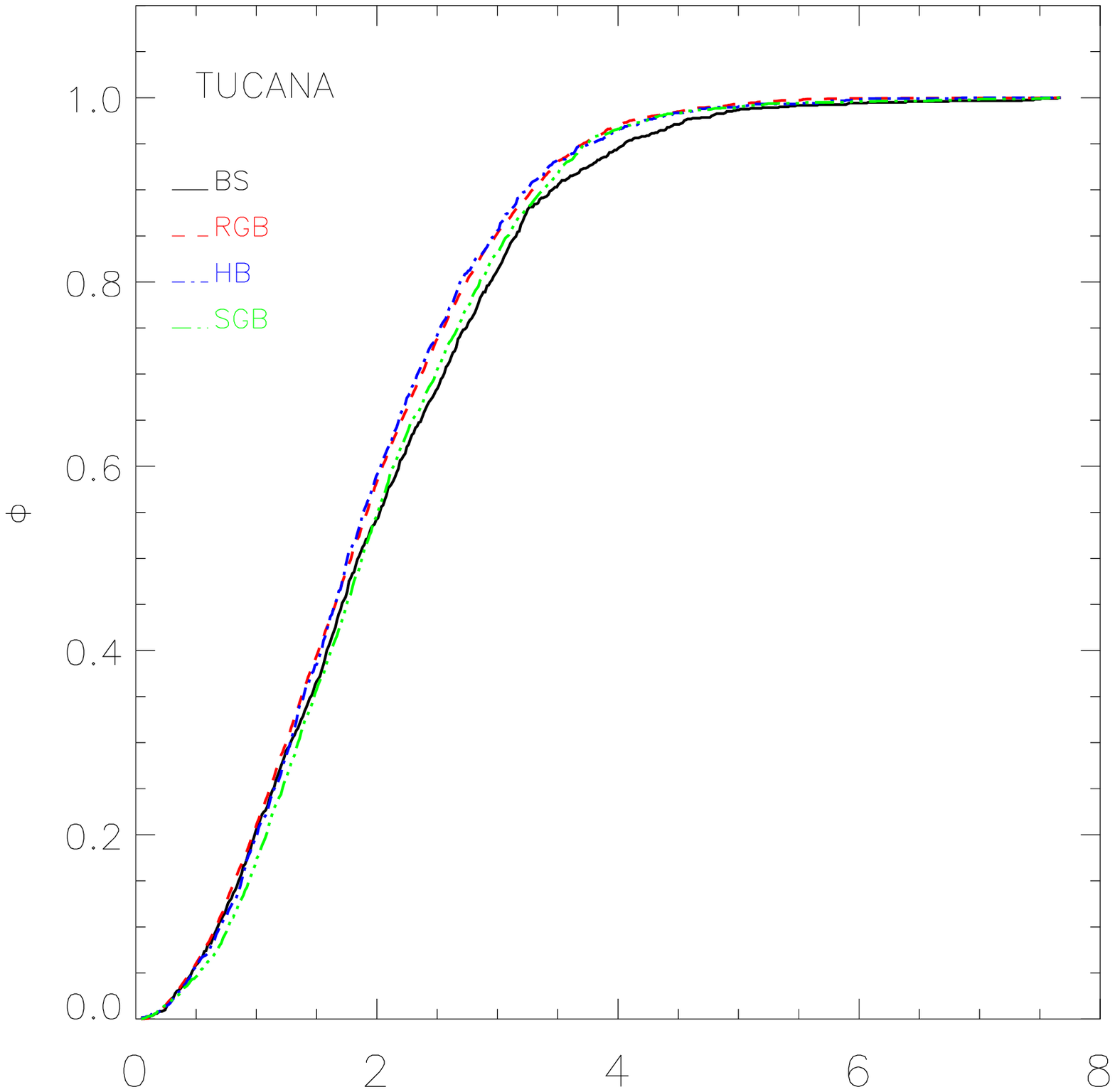}}
 \vskip 0.4cm
 \caption{Normalized cumulative distributions for the samples of BSSs (black solid line), 
 RGB (red dashed), HB (blue dot-dashed), and SGB (green) stars for Cetus (left panel) 
 and Tucana (right).}
\vspace{0.2cm}
\label{fig:cumul}
\end{figure*}


	\subsection{Luminosity function}\label{sec:lumfun}

The luminosity function of blue plume stars is a powerful diagnostic to both infer
information on their nature and, in case of a BSS population, to distinguish between the
two main formation mechanisms.
In fact, if these stars belong to an intermediate-age population, we expect
to find them more centrally concentrated (see next section) and also that
the brightest (youngest) are preferentially located to the innermost regions. On the
other hand,
it has been found that in GGCs the BSS luminosity function changes 
as a function of radial distance \citep{monkman06}, in the sense that the brightest BSS 
stars are preferentially located in the central regions. Since there is no doubt
that in GC we only have BSSs, with no contamination from intermediate-age stars, this 
occurrence is explained with a combination of two different formation mechanisms.
Collisions, which are efficient in the inner regions of GGC, tend 
to create bluer and brighter BSSs \citep{bailyn95, bailyn95R}. In the outer parts of 
the GGCs, the BSSs are predominantly created from primordial binaries. In dSph galaxies 
the formation of COL-BSS is highly disfavored due to the very low stellar density, 
so that we do not expect to observe a similar behavior \citep{mapelli07, 
momany07}. Therefore, in case a central concentration of the brightest blue plume 
stars were observed, this would point to the presence of a genuine intermediate-age population.

Figure~\ref{fig:lumfun_rad} shows the BSS candidates luminosity function for Cetus (left panel) 
and Tucana (right panel). The plot shows the luminosity function of 
stars inside (dotted line) and outside (dashed line) a distance of 1.5 times the core 
radius, $r_c$. Due to the different number of stars and area covered by the two 
samples, the curves were normalized to their area. The shape of the luminosity
function does not change significantly as a function of radius, and similarly we do not
find any evidence for an obvious shift in the magnitude of the peak.

We find that the fraction of the brightest BSSs ($BS^{br}$, i.e., within 0.5 mag of 
the brightest), is also constant with radius. For the inner and outer regions we found 
$N(BS^{br})/N(BS)=0.12\pm0.02$ and $0.12\pm0.03$ in the case of Cetus, and 
$N(BS^{br})/N(BS)=0.12\pm0.03$ and $0.13\pm0.02$ for Tucana. Note that the errors 
are solely Poisson statistical errors.

Thus, we find no evidence for segregation in either galaxy, supporting the absence
of centrally concentrated bright stars, and supporting
that the population of BSSs in Cetus and Tucana is formed as a result of the 
evolution of binary systems, even in the innermost regions. 

	\subsection{Radial distribution}\label{sec:rad}

This last result can be further verified studying the radial distribution of candidate
BSSs compared 
to that of other stellar tracers. It has been found in many GGCs that the radial profile
of BSSs has a central cusp \citep[c.f.,][]{bailyn95R}, which falls to a minimum 
\citep[the 'zone of avoidance'][]{mapelli04}, and then increases with radius. 
This phenomenon has again been interpreted with the different formation 
mechanisms at work: the COL-BSSs are more centrally concentrated than MT-BSS 
\citep{bailyn95}, which instead dominate in the outskirts of clusters. If COL-BSSs are
not present in dSphs, we expect no central concentration, as verified by \citet{mapelli07}
and \citet{mapelli09} for Draco, Ursa Minor and Sculptor.

Figure \ref{fig:grad_rad} presents the ratios $N_{BS\_HB} = N_{BS}/N_{HB}$,
$N_{BS\_RGB} = N_{BS}/N_{RGB}$, and $N_{BS\_SGB} = N_{BS}/N_{SGB}$ for Cetus (left) and Tucana
(right), as a function of the galactocentric distance expressed in units of
core radius. We divided the complete catalog into five (Cetus) and six (Tucana) radial bins,
containing the same total number of stars, according to the position of stars in elliptical
annuli. The points in Figure \ref{fig:grad_rad} show the ratio in each elliptical 
radius, plotted as a function of the mean galactocentric distance of each annulus.

Figure \ref{fig:grad_rad} clearly shows the absence of strong central concentrations 
of BSSs in both galaxies. We tested the hypothesis that the ratios of the various 
samples are consistent with a flat distribution by comparing the ratios to the mean of 
the observed distribution. In the case of Cetus, we estimated non-reduced $\chi^2$ of 
6.6, 7.3, 9.3 for $N_{BS\_RGB}$, $N_{BS\_SGB}$, and $N_{BS\_HB}$, respectively. 
The null hypothesis therefore has corresponding probabilities of 0.15, 0.12, and 0.05. 
This suggests that $N_{BS\_RGB}$ and $N_{BS\_SGB}$ are consistent with a flat 
distribution, while $N_{BS\_HB}$ is only marginally consistent.
In the case of Tucana, we derive $\chi^2$ values of 14.8, 10.6, 12.8 ($N_{BS\_RGB}$, 
$N_{BS\_SGB}$, $N_{BS\_HB}$), corresponding to probabilities of 0.01, 0.06, and 0.03. 
This means that $N_{BS\_RGB}$ and $N_{BS\_HB}$ are not consistent with a flat 
distribution, while $N_{BS\_SGB}$ is only marginally consistent.

Although $N_{BS\_HB}$ has larger uncertainties than the other ratios,
it is interesting to note that, qualitatively, $N_{BS\_HB}$ shows larger fluctuations
than the others. Moreover, in the range $0 < r_c < 2$ the trend for the two galaxies 
is opposite. Cetus presents an initial increase followed by a drop, while in Tucana 
an initial decrease within the core radius is followed by the increase in the ratio of 
BSSs relative to HB stars. At large distance, there is mild evidence of an increase
of $N_{BS\_HB}$, especially for Tucana where the field covers a large area in terms of $r_c$.
This is in agreement with the predictions by \citet{mapelli06}.

Another way to represent the radial distribution of different samples is to use the
cumulative distributions, shown in Figure \ref{fig:cumul}  for the BSS, RGB, SGB,
and HB stars. A Kolmogorov-Smirnov test gives  
probabilities of 0.861 (0.004), 0.48 (0.02), and 0.58 (0.28) for Cetus (Tucana) that the BSSs
and the RGB/HB/SGB samples are drawn from the same parent population.

However, it is interesting to note that, starting from 3 $r_c$, the BSS cumulative 
distribution is systematically lower, in both galaxies, than the others. This 
indicates higher BSS frequencies at larger distances. 

	\subsection{Frequency {\sl versus} M$_V$}\label{sec:fm}


\begin{figure}
\centering
\includegraphics[width=8.5cm]{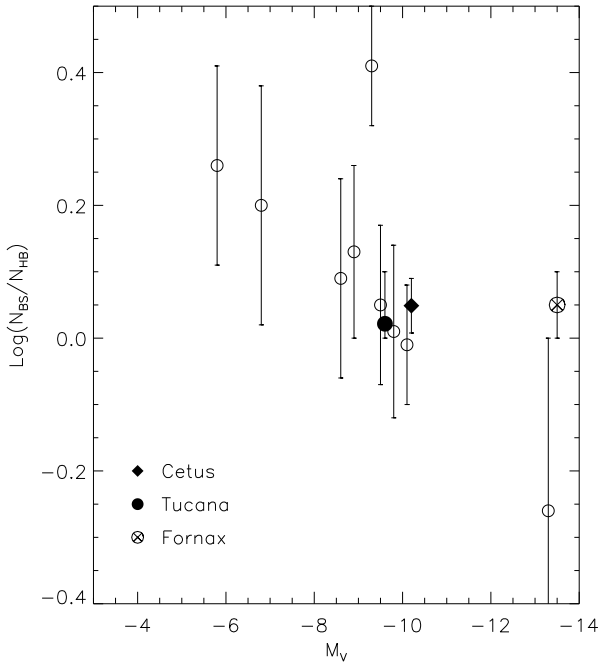}
\caption{The logarithmic frequency of BSSs in Tucana (black circle) and Cetus 
(black diamond) as a function of the absolute visual magnitude. We also show
the data corresponding to other LG dwarfs as obtained from \citet{momany07}, and
Fornax (crossed circle, Del Pino et al, 2011, in prep.).}
\label{fig:fm}
\end{figure}


\citet{momany07} found that a linear correlation exists between the logarithmic 
frequency of BSSs (F$_{BS}$, normalized to the number of HB stars) and the absolute 
visual magnitude of nearby LG dwarfs. Interestingly enough, they found that this 
correlation is verified over a large range of {\bf magnitudes}, and that its slope
is significantly different, i.e., less steep, than a similar relation found 
by \citet{piotto04} for GGCs. Presently, an underlying physical mechanism for this trend has
not been identified.  Based on the selections presented
in Figure \ref{fig:samples}, we estimated that F$_{BS}$ is equal to 0.05$\pm$0.05
and 0.02$\pm$0.04 for Cetus and Tucana, respectively. 

In Figure~\ref{fig:fm}, these values are plotted together with data from other galaxies 
given in Figure~2 of \citet{momany07}. We adopted the absolute $M_V$ magnitude
from \citet{mcconnachie06} for Cetus ($-10.1$ mag) and from \citet{saviane96} for Tucana 
($-9.6$ mag). Fig.~\ref{fig:fm} shows that Tucana and Cetus are consistent with
the relationship derived from other LG dwarfs. 
Using the analytic relation for $F_{BS}$ from \citet{momany07}, we obtain $-0.01\pm$0.08
and 0.03$\pm$0.08 for Cetus and Tucana, respectively, again, showing very good agreement
with the observed values.

If we accept that the $F_{BS} - M_V$ relation is significant, then this is an independent
indication that the we are dealing with a genuine BSS population. If the blue plume stars 
were due to fluctuations in the SFH producing younger stars, one would not expect any 
relationship between $F_{BS}$ and $M_V$.  On the other hand, if blue plume stars are BSSs, 
they essentialy belong to the same population generating the HB stars. As already noted by 
\citet{momany07}, the trend is similar in the case of globular clusters, but with 
significantly shallower slope. The existence of an anti-correlation with the luminosity
({\bf the higher the luminosity, the lower the BSS frequency}), and hence with the mass of 
the cluster, might suggest that in more massive globular clusters the mechanisms
affecting the binary systems are more efficient then in less massive ones, and this
cause a deficiency of BSS with respect of HB stars \citep[see][]{piotto04}. 
In dSph, the shallower slope might indicate that the same mechanisms are still at work, 
but are less efficient due to the lower density of dwarf galaxies.

On the basis of present results, it appears evident that if the reliability of $F_{BS} -
M_V$ relation was confirmed by increasing the number of LG dwarfs in the sample, it
could represent a valuable tool for discriminating stellar systems hosting truly young MS
stars from those with a significant population of old, genuine BSSs.  For example, a
preliminary analysis of a deep CMD of Fornax dSph (Del Pino et al.~2011, in preparation, 
 also shown in Figure~\ref{fig:fm} as a crossed-circle symbol)
suggests that the F$_{BS}$ estimate for this galaxy (F$_{BS} > 0.05\pm0.05$) is significantly
higher than  expected from its absolute magnitude (F$_{BS} = -0.23\pm0.08$). This is in
agreement with the conclusion presented in \citet{mapelli09}, and with the fact that 
Fornax hosts a population as young as 100~Myr \citep[see also][]{stetson98}. 


\begin{figure*}
\resizebox{8truecm}{7truecm}{\includegraphics{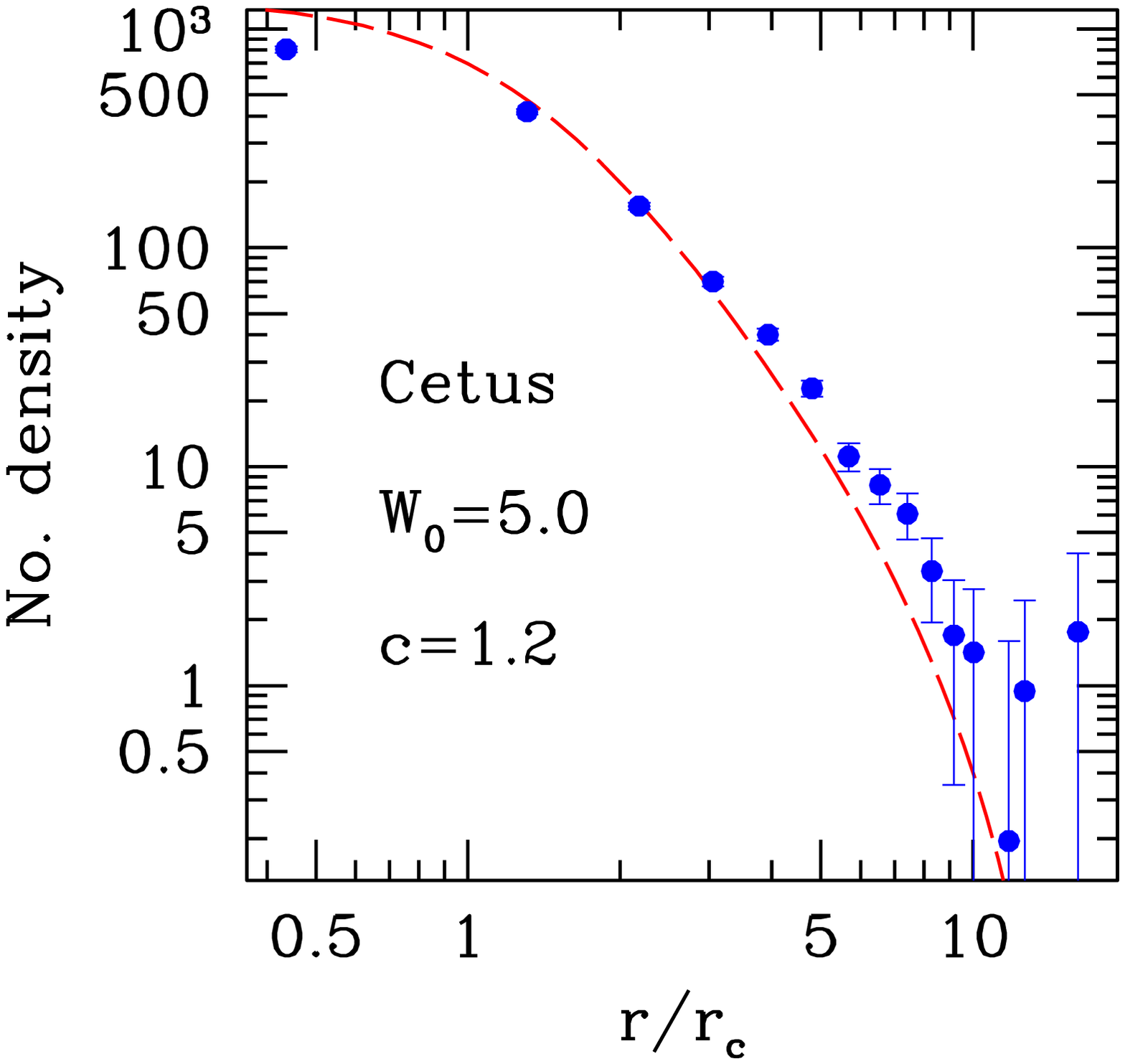}}
\resizebox{8truecm}{7truecm}{\includegraphics{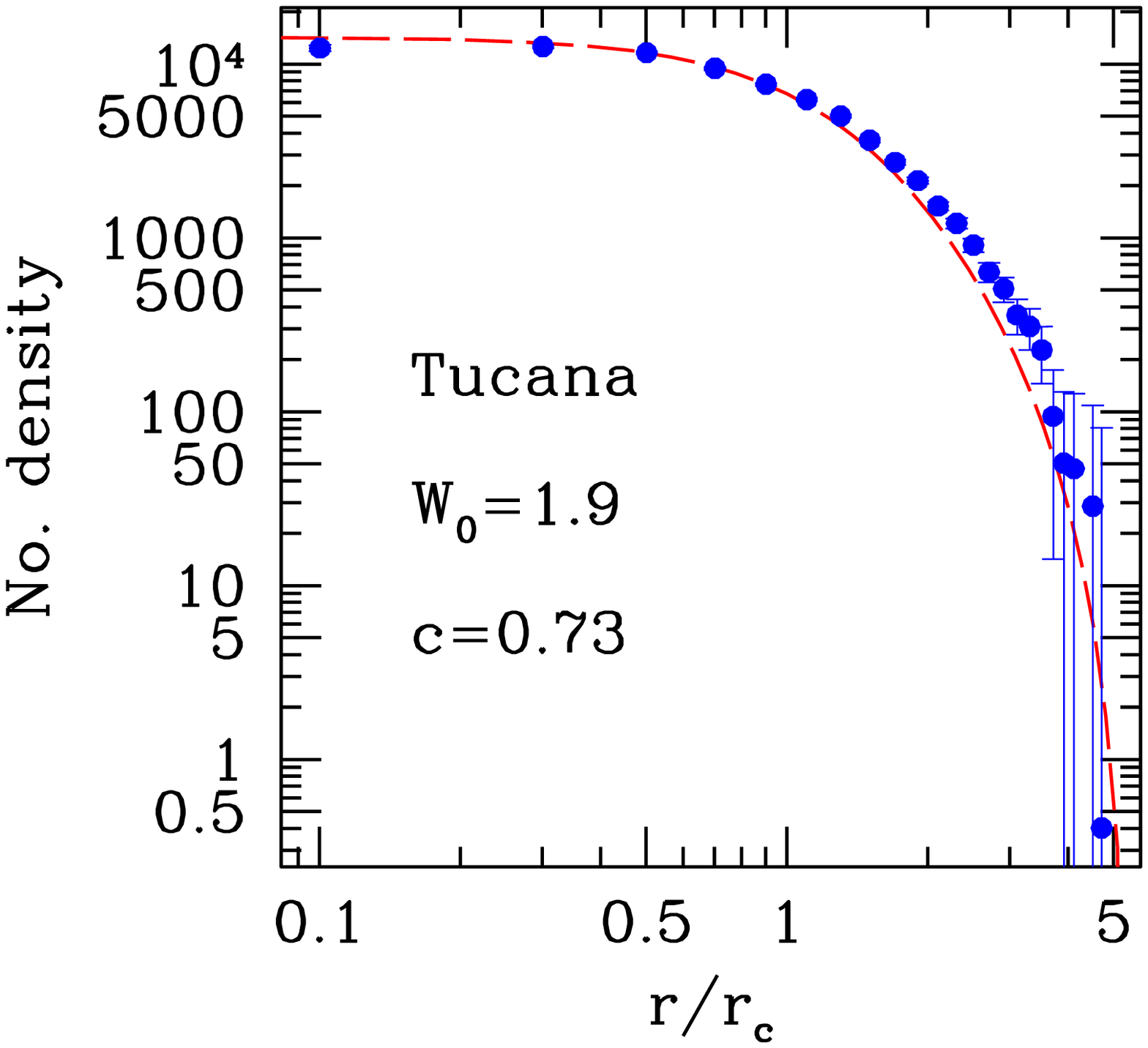}}
 \caption{Simulated stellar radial profile (dashed line, red on the web) for Cetus 
 (left-hand panel) and Tucana (right-hand panel), compared with observations (filled circles, 
 blue on the web). The number density is given in stars per square arcmin. Data points for
 Cetus are based on deep Subaru data (Bernard et al.\ 2011, in prep.), while are taken from 
 \citet{bernard09} for Tucana.}\label{fig:simprofile} 
\end{figure*}



\begin{figure*}
\resizebox{8truecm}{7truecm}{\includegraphics{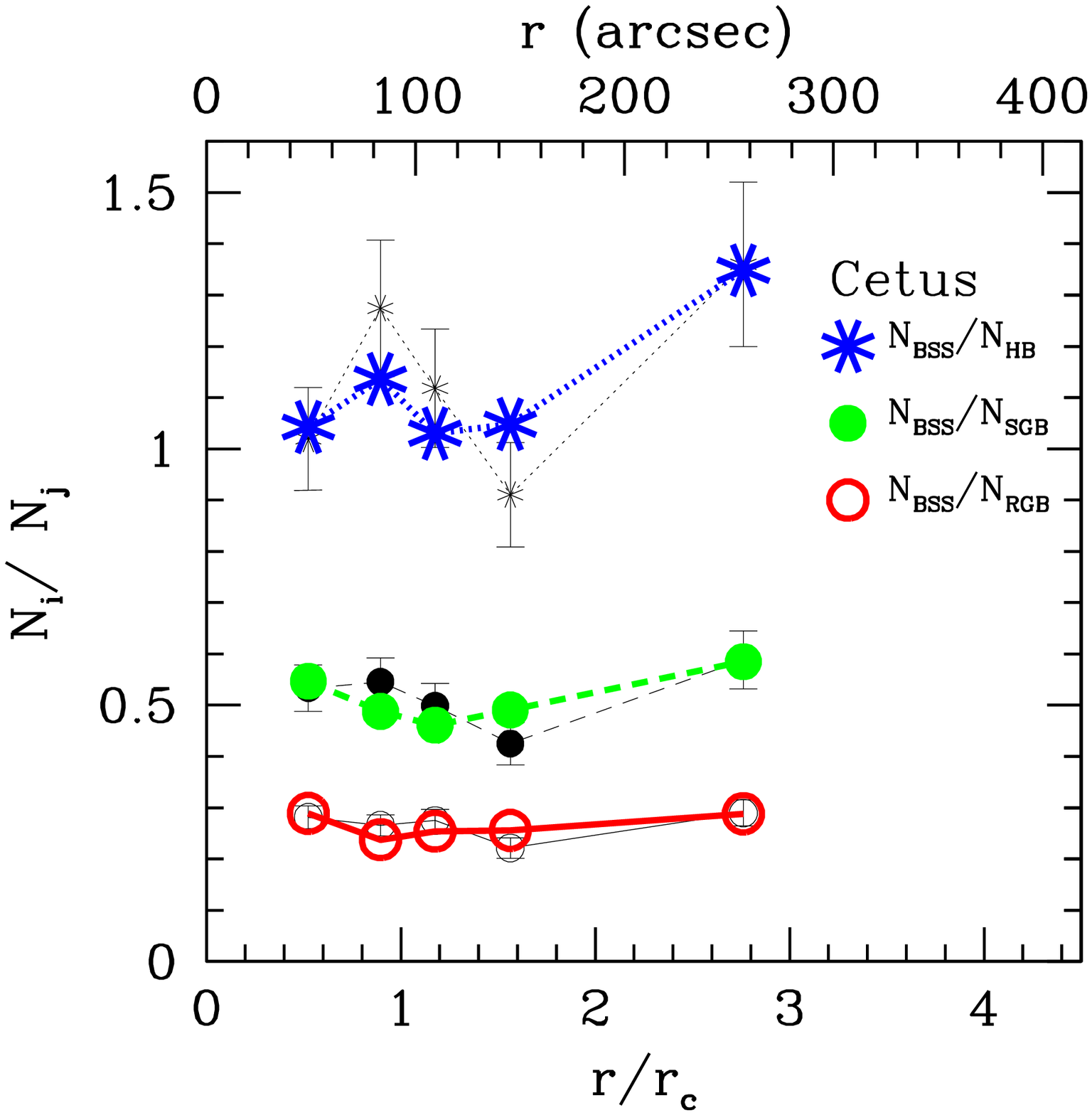}}
\resizebox{8truecm}{7truecm}{\includegraphics{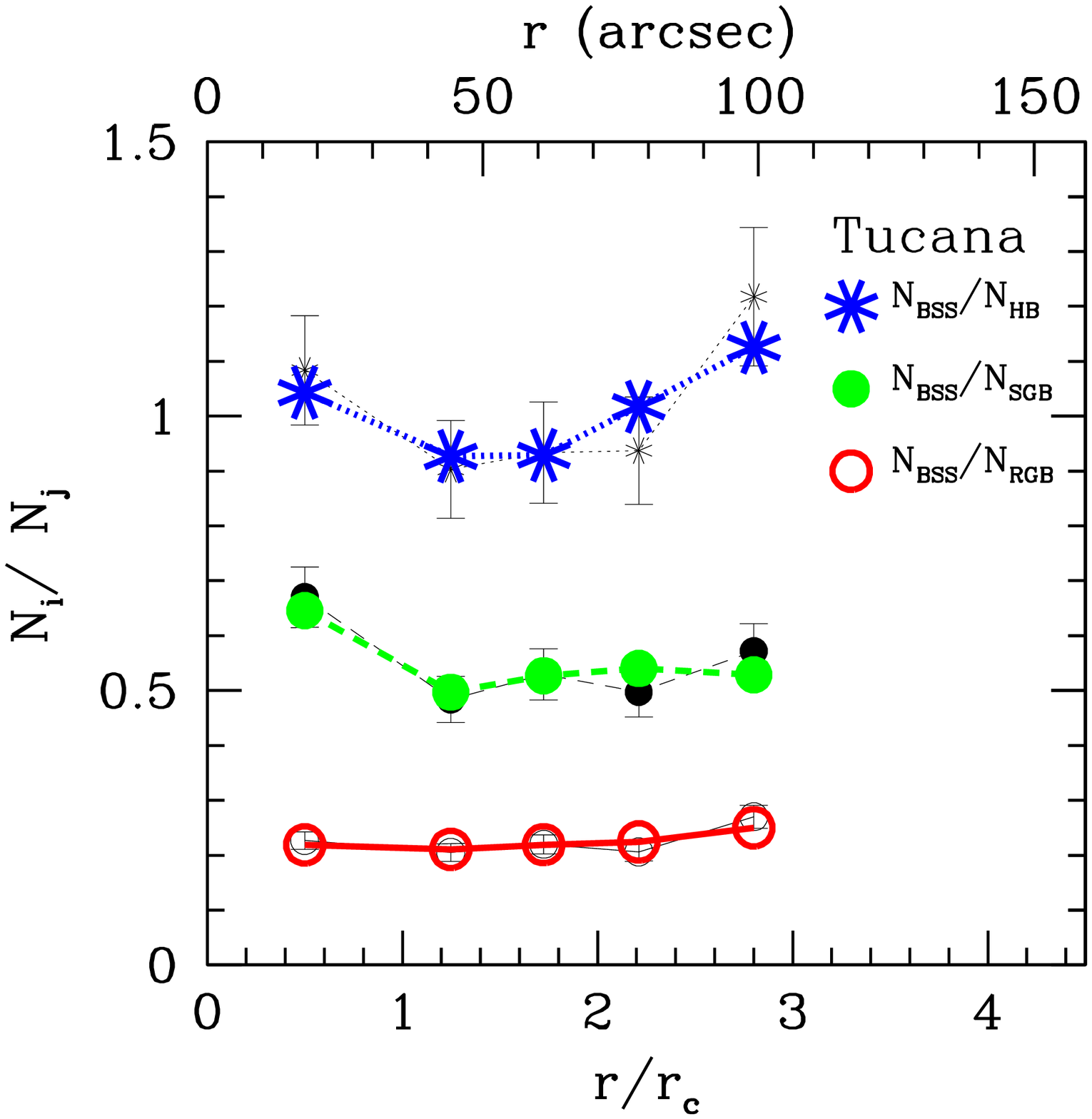}}
 \caption{Simulated radial distribution of $N_{\rm BS\_RGB}$ 
 , $N_{\rm BS\_SGB}$ and $N_{\rm BS\_HB}$ 
 in Cetus (left-hand panel) and in Tucana (right-hand panel). Thin 
 lines indicate the observed radial distributions,  for comparison with the 
 best match simulations (runs C3 and T3 for Cetus and Tucana, respectively). 
 Error bars account for Poisson uncertainties.}
\label{fig:dyn} 
\end{figure*}


\section{Dynamical simulations}\label{sec:simul}

Further clues about the origin of BSSs can be inferred from their dynamical 
evolution \citep[see, e.g., ][]{sigurdsson94}. We thus perform a 
wide sample of dynamical simulations, following the evolution of MT-BSS and COL-BSS 
populations in the potential of Cetus and Tucana. 

	\subsection{Method}\label{sec:simulmeth}

We  use  the code BEV \citep{sigurdsson95}, in the upgraded version described by
\citet{mapelli04, mapelli06}. The same code has been used to model BSSs in Draco, Ursa Minor
\citep{mapelli07}, Sculptor and Fornax \citep{mapelli09}. BEV integrates the dynamical
evolution of BSSs under the influence of the host-galaxy potential, dynamical friction, 
distant encounters with other stars, and three-body encounters (which are negligible for
dSphs). The potential of the host galaxy is assumed to be time-independent and is
represented by a multi-mass King model. The classes of mass are the same as in
\citet{mapelli07}, and the TO mass is assumed to be 0.8 $M_\odot{}$ (see Figure
\ref{fig:track}) for both Cetus and Tucana. To calculate the potential, we input the number
density of stars in the core ($n_c=0.03$ and $=0.12$ stars pc$^{-3}$ for Cetus and Tucana,
respectively) the velocity dispersion in the core (we assume $\sigma_{c}=10$ km s$^{-1}$ for
both Cetus and Tucana\footnotemark[15]\footnotetext[15]{The observed velocity dispersion,
$v_{obs}$, is higher ($\sim{}17$ km s$^{-1}$) for both Cetus \citep{lewis07} and Tucana
\citep{fraternali09}, but there is evidence that a significant part of it might be due to
rotation.}),  and we modify the central adimensional potential $W_0$ 
\citep[defined in][]{sigurdsson95}, equating the simulated concentration and radial density profile with
the observed values. Fig.~\ref{fig:simprofile} shows the resulting models for Cetus (obtained
for $W_0=5$ and $c=1.20$) and for Tucana (obtained for $W_0=1.9$ and $c=0.73$).  In the case
of Tucana, the simulated profile and the concentration of the model match the observations
reported in \citet{bernard09}. For the concentration of Cetus, we adopt $c=1.2$, which best
matches our data. We note that the model for Cetus has a relatively high value of $W_0$, with 
respect to other dSphs (generally, $W_0\ge{}6$ for globular clusters and $W_0\lesssim{}3$ 
for dSphs).

In our model, BSSs are generated with a given position, velocity, mass and lifetime. The
fundamental differences between COL-BSSs and MT-BSSs are the following: 

(i) The distribution of
initial positions: the initial positions of MT-BSSs are randomly chosen according to a
probability distribution homogeneous in radial distance from the center. This means that
MT-BSSs are initially distributed according to an isothermal sphere, since MT-BSSs are expected to
follow the same distribution as the  primordial binaries \citep[see ][]{mapelli04, mapelli06}.
(The spherical nature of the dSphs is not conclusively established, but is chosen as 
a reasonable starting point.)
 The minimum and the maximum value of the distribution of initial  radial distances of MT-BSSs, $r_{\rm
min,\,{}MT}$ and $r_{\rm max,\,{}MT}$ (both defined as three-dimensional quantities), have been tuned in
order  to find the best-fitting simulation. On the other hand, COL-BSSs are generated only inside $r_c$ (i.e., $r_{\rm
min,\,{}COL}=0$ and $r_{\rm max,\,{}COL}=r_{\rm c}$ in all the simulations),
the only region were collisions might occur, although collisions are highly unlikely in
dSphs (the probability of a three-body interaction in the core of Cetus and Tucana is
$\gtrsim{}10^5$ times lower than in a moderately dense globular cluster). The initial
distribution of COL-BSSs is not necessarily homogeneous in the radial distance, as they are not
expected to follow the same distribution as binaries (in our simulations, we assume a
constant probability distribution between the center of the cluster and $r_c$).

(ii) The distribution of initial velocities: the initial velocities of MT-BSSs are drawn from the
equilibrium distributions of dSph stars with the same masses as the MT-BSSs. COL-BSSs are
assumed to be born with a natal kick \citep[see][]{sigurdsson94}. We adopt a kick velocity,
$v_{\rm kick}$, $=1-2\,{}\sigma{}_c$. For kick velocities higher than $2\,{}\sigma{}_c$ most of
COL-BSSs are ejected from the dSph, whereas for kick velocities lower than $1\,{}\sigma{}_c$,
COL-BSSs are almost indistinguishable from MT-BSSs from a dynamical point of view.

In most of the runs, the masses of the BSSs are assumed to be $m_{\rm BS}$ = 1.3 M$_\odot{}$. We
also calculated simulations with $m_{\rm BS}$ = 0.9, 1.1 M$_\odot{}$, but the results are
substantially unchanged. Each BSS is evolved for a time $t$, randomly selected from a
homogeneous distribution between $t=0$ and $t=t_{\rm life}$. The parameter $t_{\rm life}$ is
the lifetime of BSSs \citep[see ][]{mapelli04}. We calculated simulations with $t_{\rm life}=$ 
2 and 4 Gyr (4 Gyr is the age inferred from the derived SFH).

	\subsection{Comparison with observations}\label{sec:simulcompa} 

Tables \ref{tab:tab02} and \ref{tab:tab03} show the parameters of the 
simulations and report also the fraction $\eta{}$ of COL-BSSs present in each run 
($\eta{}\equiv{}N_{\rm COL-BSS}/(N_{\rm COL-BSS }+N_{\rm MT-BSS})$, where $N_{\rm COL-BSS}$ and 
$N_{\rm MT-BSS}$ are the number of COL-BSSs and of MT-BSSs, respectively). The results of the 
$\chi^2$ analysis are also included. For each run (we made 38 runs for Cetus and 37 for Tucana)
10000 BSSs have been simulated. We show three values of the non-reduced $\chi^2$, referring
to $N_{\rm BS\_RGB}$ ($\chi{}^2_{\rm RGB}$, column 8 of Tables \ref{tab:tab02} and
\ref{tab:tab03}), $N_{\rm BS\_SGB}$ ($\chi{}^2_{\rm SGB}$, column 9) and $N_{\rm BS\_HB}$
($\chi{}^2_{\rm HB}$, column 10). Fig.~\ref{fig:dyn} shows the radial distributions of $N_{\rm
BS\_RGB}$, $N_{\rm BS\_SGB}$, and $N_{\rm BS\_HB}$ obtained for the best-matching simulations
(run C3 and  run T7 for Cetus and  Tucana, respectively), compared with the observed
distributions (the same as in Fig.~\ref{fig:grad_rad}). We stress that only the
radial distribution of $N_{\rm BSS}$ comes from the simulations, and it is compared with the
observed radial distribution of $N_{\rm RGB}$, $N_{\rm HB}$ and $N_{\rm SGB}$. In fact, the 
adopted simulation method does not allow to evolve these specific stellar types, but only stars
in a given class of mass (e.g., Mapelli et al. 2004). We also remind that the underlying potential 
of the dSph is analytic and time-independent\footnotemark[16]\footnotetext[16]{A fully 
N-body simulation, which may account for
the dynamical and stellar evolution of all the aforementioned populations, would give a
much more realistic description of the system, but the computational time
for running a complete grid of such simulations is prohibitive even for star
clusters, let alone for dwarf spheroidals.}. The best-matching simulations
reproduce well the observed radial distributions of $N_{\rm BS\_RGB}$, $N_{\rm
BS\_SGB}$, and $N_{\rm BS\_HB}$. In fact, the values of the minimum non-reduced $\chi{}^2$
are $\chi{}^2_{\rm RGB}\lesssim{}6$, $\chi{}^2_{\rm SGB}\lesssim{}5$ and $\chi{}^2_{\rm
SGB}\lesssim{}4$ for Cetus (run C3) and $\chi{}^2_{\rm RGB}\lesssim{}3$, $\chi{}^2_{\rm
SGB}\lesssim{}2$ and $\chi{}^2_{\rm SGB}\lesssim{}2$ for Tucana (runs T5, T7, T14, T15, and
T16). 

\begin{figure}
\resizebox{8truecm}{8truecm}{\includegraphics{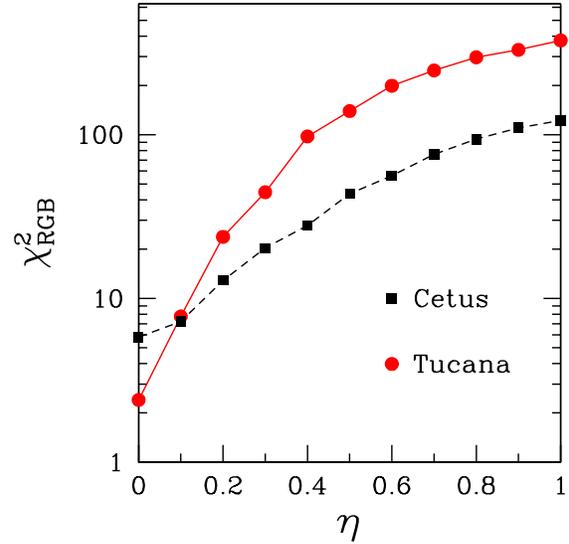}}
\vspace{0.5cm}
 \caption{$\chi{}^2_{\rm RGB}$ as a function of the fraction of COL-BSSs ($\eta{}$) for the 
simulations of Cetus (squares connected by the dashed line) and Tucana (circles connected by 
the solid line). For each run,
only eta was changed, while the other parameters are the same of the best models (C3 and
T7 for Cetus and Tucana respectively) and were held constant.
}
 \label{fig:chisq}
\end{figure}

Fig~\ref{fig:chisq} shows the behavior of $\chi{}^2_{\rm RGB}$ (that of $\chi{}^2_{\rm SGB}$ 
and of $\chi{}^2_{\rm HB}$ is very similar, see Tables \ref{tab:tab02} and \ref{tab:tab03}) as a 
function of the fraction of COL-BSSs ($\eta{}$). 
As it was reasonable to expect, adding a population of COL-BSSs significantly reduces the agreement 
between data and model: for both Cetus and Tucana,
only simulations with $\eta{}<0.2$ give an acceptable $\chi{}^2$.


\begin{figure}
\resizebox{8truecm}{8truecm}{\includegraphics{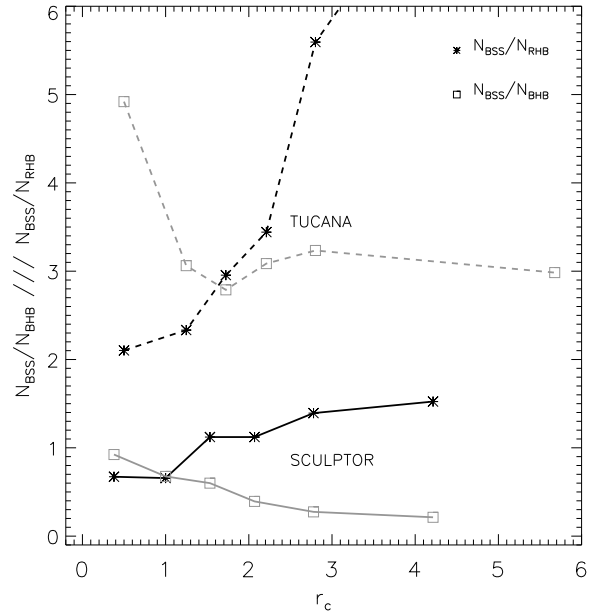}}
\vspace{0.5cm}
 \caption{The trend with the radial distance of the stellar population ratios 
 $N_{BSS}/N_{BHB}$ and $N_{BSS}/N_{RHB}$ for both Tucana and Sculptor
 (see text for more details).}
 \label{fig:scltuc}
\end{figure}


The dependence of the results on the BSSs mass (in the range allowed by the observations) is
negligible. This is in agreement with the findings of previous papers \citep{mapelli07,
mapelli09} for BSSs in Draco, Ursa Minor, and Sculptor. The dependence on the lifetime of BSSs
is negligible for Cetus (where the runs with $t_{\rm life}=2$ Gyr are almost equivalent to
the runs with  $t_{\rm life}=4$ Gyr) and slightly more important for Tucana (where the runs
with $t_{\rm life}=2$ Gyr have generally lower values of $\chi{}^2$ with respect to the runs
with  $t_{\rm life}=4$ Gyr, but the difference is not statistically significant). Given the
range of masses expected for the BSS, suggested by the comparison with stellar tracks,
we assume 4 Gyr as fiducial value.

In the case of Cetus, for the best-matching runs $r_{\rm min}\sim{}0.0$ and $r_{\rm
max}\sim{}5\,{}r_c$, indicating that primordial binaries were initially distributed ranging
from the center of the dSph out to (at least) $5\,{}r_c$. The tidal radius of Cetus (derived
from our data) is 24$\arcmin$, i.e., $\sim{}15\,{}r_c$, much larger than  $r_{\rm max}$.
However, this difference likely does not mean that BSSs do not form beyond $5\,{}r_c$, but is
a consequence of the fact that we do not have data beyond $\approx{}4\,{}r_c$ (furthermore,
the coverage of Cetus is not complete even inside $4\,{}r_c$). In the case of Tucana, for
which data are available almost up to the tidal radius, the best-matching value of $r_{\rm
max}$ is equal to the tidal radius. Interestingly, the best-matching values of $r_{\rm min}$
for Tucana are all larger than one core radius.

In summary, the results of dynamical simulations show that it is possible to explain the
observed distribution of BSS candidates, in both Cetus and Tucana, with a MT-BSS population. 
Any population of COL-BSSs more numerous than the 10\% of the total number of BSSs reduces the
agreement between data and simulations. This result is not surprising, as basic dynamical 
calculations tend to exclude the existence of COL-BSSs in dSphs, but it is important to 
point out that the simulations can almost rule out the unphysical scenario where $>20$\% of 
BSSs are COL-BSSs in both Cetus and Tucana.



\begin{deluxetable*}{cccccccccc}
\tabletypesize{\scriptsize}
\tablewidth{0pt}
\tablecaption{Parameters and values of $\chi{}^2$ for the simulations of Cetus. \label{tab:tab02}}
\tablehead{
\colhead{Case $^{\rm a}$} & \colhead{$r_{\rm min,\,{}MT}/r_{\rm c}$ $^{\rm b}$} & \colhead{$r_{\rm max,\,{}MT}/r_{\rm c}$ $^{\rm b}$} & \colhead{$m_{\rm BSS}/{M_{\odot}}$ $^{\rm c}$} & \colhead{$t_{\rm life}/$ Gyr $^{\rm d}$} & \colhead{$v_{\rm kick}/\sigma{}_{\rm c}$ $^{\rm e}$} & \colhead{$\eta{}$ $^{\rm f}$} & \colhead{$\chi{}^2_{\rm RGB}$ $^{\rm g}$} & \colhead{$\chi{}^2_{\rm SGB}$ $^{\rm g}$} & \colhead{$\chi{}^2_{\rm HB}$ $^{\rm g}$} }
\startdata
C1    & 0.0                     & 5.0                     & 0.9                          & 4                   &     --  & 0                         &  7.7    &   6.5   &   4.7   \\
C2    & 0.0                     & 5.0                     & 1.1                          & 4                   &     --  & 0                         &  9.5    &   8.1   &   5.8   \\
{\bf C3}    & {\bf 0.0}                     & {\bf 5.0}                     & {\bf 1.3}                          & {\bf 4}                   &     {\bf --}  & {\bf 0}                         &  {\bf 5.8}    &   {\bf 4.9}   &   {\bf 3.5}   \\
C4    & 0.1                     & 5.0                     & 1.3                          & 4                   &     --  & 0                         &  8.3    &   7.0   &   5.0   \\
C5    & 0.2                     & 5.0                     & 1.3                          & 4                   &     --  & 0                        &  7.3    &   6.2   &   4.5   \\
C6    & 0.3                     & 5.0                     & 1.3                          & 4                   &     --  & 0                        & 10      &   8.7   &   6.2   \\
C7    & 0.4                     & 5.0                     & 1.3                          & 4                   &     --  & 0                         & 14      &  12     &   8.5   \\
C8    & 0.5                     & 5.0                     & 1.3                          & 4                   &     --  & 0                         & 19      &  16     &  12     \\
C9    & 1.0                     & 5.0                     & 1.3                          & 4                   &     --  & 0                         & 51      &  42     &  31     \\
C10   & 0.0                     & 5.0                     & 0.9                          & 2                   &     --  & 0                         &  6.7    &   5.6   &   4.1   \\
C11   & 0.0                     & 5.0                     & 1.1                          & 2                   &     --  & 0                         &  7.2    &   6.1   &   4.4   \\
C12   & 0.0                     & 5.0                     & 1.3                          & 2                   &     --  & 0                         &  9.8    &   8.3   &   5.9   \\
C13   & 0.1                     & 5.0                     & 1.3                          & 2                   &     --  & 0                         &  9.0    &   7.6   &   5.4   \\
C14   & 0.2                     & 5.0                     & 1.3                          & 2                   &     --  & 0                         &  7.0    &   5.9   &   4.2   \\
C15   & 0.3                     & 5.0                     & 1.3                          & 2                   &     --  & 0                         &  7.5    &   6.3   &   4.6   \\
C16   & 0.4                     & 5.0                     & 1.3                          & 2                   &     --  & 0                         & 12      &  10     &   7.5   \\
C17   & 0.5                     & 5.0                     & 1.3                          & 2                   &     --  & 0                         & 20      &  17     &  12     \\
C18   & 1.0                     & 5.0                     & 1.3                          & 2                   &     --  & 0                         & 50      &  41     &  30     \\
C19   & 0.0                     & 4.0                     & 1.3                          & 4                   &     --  & 0                         & 10      &   8.7   &   6.2   \\
C20   & 0.2                     & 4.0                     & 1.3                          & 4                   &     --  & 0                         &  9.4    &   7.8   &   5.5   \\
C21   & 0.5                     & 4.0                     & 1.3                          & 4                   &     --  & 0                         &  7.9    &   6.7   &   4.8   \\
C22   & 1.0                     & 4.0                     & 1.3                          & 4                   &     --  & 0                         & 41      &  34     &   25    \\
C23   & 0.0                     & 6.0                     & 1.3                          & 4                   &     --  & 0                         & 13      &  11     &    7.7  \\
C24   & 0.2                     & 6.0                     & 1.3                          & 4                   &     --  & 0                         & 16      &  13     &    9.1  \\
C25   & 0.5                     & 6.0                     & 1.3                          & 4                   &     --  & 0                         & 26      &  22     &   16    \\
C26   & 1.0                     & 6.0                     & 1.3                          & 4                   &     --  & 0                         & 77      &  64     &   46    \\
C27   & 0.0                     & 5.0                     & 1.3                          & 4                   &     1  &  0.1                       &  7.2    &   6.0   &    4.3  \\
C28   & 0.0                     & 5.0                     & 1.3                          & 4                   &     1  &  0.2                       & 13      &  11     &    7.8  \\
C29   & 0.0                     & 5.0                     & 1.3                          & 4                   &     1  &  0.3                       & 20      &  17     &   12    \\
C30   & 0.0                     & 5.0                     & 1.3                          & 4                   &     1  &  0.4                       & 28      &  23     &   16    \\
C31   & 0.0                     & 5.0                     & 1.3                          & 4                   &     1  &  0.5                       & 43      &  36     &   26    \\
C32   & 0.0                     & 5.0                     & 1.3                          & 4                   &     1  &  0.6                       & 56      &  47     &   33    \\
C33   & 0.0                     & 5.0                     & 1.3                          & 4                   &     1  &  0.7                       & 76      &  63     &   45    \\
C34   & 0.0                     & 5.0                     & 1.3                          & 4                   &     1  &  0.8                       & 94      &  78     &   56    \\
C35   & 0.0                     & 5.0                     & 1.3                          & 4                   &     1  &  0.9                       & 110     &  91     &   65    \\
C36   & --                      & --                      & 1.3                          & 4                   &     1  &  1.0                       & 122     & 101     &   72    \\
C37   & 0.0                     & 5.0                     & 1.3                          & 4                   &     2  &  0.1                       & 9.4     &  7.9    &    5.7  \\
C38   & 0.0                     & 5.0                     & 1.3                          & 4                   &     2  &  0.2                       & 14      & 12      &    8.3 \\
\enddata
\tablenotetext{a}{Identifier of the run (for each run, 10000 BSSs have been simulated).}
\tablenotetext{b}{$r_{\rm min,\,{}MT}$ and $r_{\rm max,\,{}MT}$ are the minimum and the maximum three-dimensional radius within which MT-BSSs are generated in the simulations. The minimum and the maximum three-dimensional radius within which COL-BSSs are generated in the simulations are not listed in the Table, because they are the same in all the simulations (i.e., $r_{\rm min,\,{}COL}=0$ and $r_{\rm max,\,{}COL}=r_{\rm c}$, respectively).}
\tablenotetext{c}{Mass of a simulated  BSS. In the simulations, $m_{\rm BSS}$ is the same for MT-BSSs and COL-BSSs.}
\tablenotetext{d}{Lifetime of a simulated BSS. In the simulations, $t_{\rm life}$ is the same for MT-BSSs and COL-BSSs.}
\tablenotetext{e}{Kick velocity for COL-BSSs.} 
\tablenotetext{f}{Fraction of simulated COL-BSSs with respect to the total population of BSSs.} 
\tablenotetext{g}{$\chi{}^2_{\rm RGB}$, $\chi{}^2_{\rm SGB}$ and $\chi{}^2_{\rm HB}$ are the non-reduced $\chi{}^2$ of the simulated $N_{\rm BS\_RGB}$, $N_{\rm BS\_SGB}$ and $N_{\rm BS\_HB}$.}
\end{deluxetable*}

\begin{deluxetable*}{cccccccccc}
\tabletypesize{\scriptsize}
\tablewidth{0pt}
\tablecaption{Parameters and values of $\chi{}^2$ for the simulations of Tucana. \label{tab:tab03}}
\tablehead{
\colhead{Case $^{\rm a}$} & \colhead{$r_{\rm min,\,{}MT}/r_{\rm c}$ $^{\rm b}$} & \colhead{$r_{\rm max,\,{}MT}/r_{\rm c}$ $^{\rm b}$} & \colhead{$m_{\rm BSS}/{M_{\odot}}$ $^{\rm c}$} & \colhead{$t_{\rm life}/$ Gyr $^{\rm d}$} & \colhead{$v_{\rm kick}/\sigma{}_{\rm c}$ $^{\rm e}$} & \colhead{$\eta{}$ $^{\rm f}$} & \colhead{$\chi{}^2_{\rm RGB}$ $^{\rm g}$} & \colhead{$\chi{}^2_{\rm SGB}$ $^{\rm g}$} & \colhead{$\chi{}^2_{\rm HB}$ $^{\rm g}$} }
\startdata
T1    & 0.0                     & 5.4                     & 1.3                          & 4                   &     0 & --         & 101     &  78   &  60   \\
T2    & 0.5                     & 5.4                     & 1.3                          & 4                   &     0 & --         &  51     &  40   &  30   \\
T3    & 1.0                     & 5.4                     & 1.3                          & 4                   &     0 & --         &  21     &  17   &  12   \\
T4    & 1.5                     & 5.4                     & 1.3                          & 4                   &     0 & --         &   5.7   &   4.6 &   3.4 \\
T5    & 1.7                     & 5.4                     & 0.9                          & 4                   &     0 & --         &   2.6   &   2.1 &   1.6 \\
T6    & 1.7                     & 5.4                     & 1.1                          & 4                   &     0 & --         &   3.2   &   2.6 &   1.9 \\
{\bf T7}    & {\bf 1.7}                     & {\bf 5.4}                     & {\bf 1.3}                          & {\bf 4}                   &     {\bf 0} & {\bf --}         &   {\bf 2.4}   &   {\bf 1.9} &   {\bf 1.5} \\
T8    & 2.0                     & 5.4                     & 1.3                          & 4                   &     0 & --         &   7.7   &   6.1 &   4.8 \\
T9    & 2.5                     & 5.4                     & 1.3                          & 4                   &     0 & --         &  18     &  14   &  11   \\

T10   & 0.0                     & 5.4                     & 1.3                          & 2                   &     0 & --         & 106     &  83   &  63   \\
T11   & 0.5                     & 5.4                     & 1.3                          & 2                   &     0 & --         &  55     &  43   &  33   \\
T12   & 1.0                     & 5.4                     & 1.3                          & 2                   &     0 & --         &  23     &  18   &  14   \\
T13   & 1.5                     & 5.4                     & 1.3                          & 2                   &     0 & --         &   2.5   &  2.1  &   1.5 \\
T14   & 1.7                     & 5.4                     & 0.9                          & 2                   &     0 & --         &   2.3   &  1.8  &   1.4 \\
T15   & 1.7                     & 5.4                     & 1.1                          & 2                   &     0 & --         &   1.9   &  1.5  &   1.1 \\
T16   & 1.7                     & 5.4                     & 1.3                          & 2                   &     0 & --         &   1.6   &  1.3  &   0.95 \\
T17   & 2.0                     & 5.4                     & 1.3                          & 2                   &     0 & --         &   4.8   &  3.7  &   2.9  \\
T18   & 2.5                     & 5.4                     & 1.3                          & 2                   &     0 & --         &  16     &  13   &   9.8 \\

T19   & 0.0                     & 4.0                     & 1.3                          & 4                   &     0 & --         & 139     & 109   &  83   \\
T20   & 0.5                     & 4.0                     & 1.3                          & 4                   &     0 & --         &  83     &  65   &  49   \\
T21   & 1.0                     & 4.0                     & 1.3                          & 4                   &     0 & --         &  41     &  32   &  24   \\
T22   & 1.5                     & 4.0                     & 1.3                          & 4                   &     0 & --         &  15     &  12   &   8.6  \\
T23   & 1.7                     & 4.0                     & 1.3                          & 4                   &     0 & --         &   5.5   &   4.5 &   3.3  \\
T24   & 2.0                     & 4.0                     & 1.3                          & 4                   &     0 & --         &   6.3   &   4.9 &   3.8  \\
T25   & 2.5                     & 4.0                     & 1.3                          & 4                   &     0 & --         &  11     &   8.7 &   6.7  \\   

T26   & 1.7                     & 5.4                     & 1.3                          & 4                   &     1 & 0.1       &   7.8   &   6.0 &   4.5  \\
T27   & 1.7                     & 5.4                     & 1.3                          & 4                   &     1 & 0.2       &  24     &  18   &    14  \\
T28   & 1.7                     & 5.4                     & 1.3                          & 4                   &     1 & 0.3       &  45     &  34   &    26  \\
T29   & 1.7                     & 5.4                     & 1.3                          & 4                   &     1 & 0.4       &  98     &  76   &    58  \\
T30   & 1.7                     & 5.4                     & 1.3                          & 4                   &     1 & 0.5       & 140     &  109  &    84  \\
T31   & 1.7                     & 5.4                     & 1.3                          & 4                   &     1 & 0.6       & 199     &  156  &   120  \\
T32   & 1.7                     & 5.4                     & 1.3                          & 4                   &     1 & 0.7       & 247     &  195  &   149  \\
T33   & 1.7                     & 5.4                     & 1.3                          & 4                   &     1 & 0.8       & 297     &  236  &   180  \\
T34   & 1.7                     & 5.4                     & 1.3                          & 4                   &     1 & 0.9       & 330     &  262   &  200  \\
T35   & --                      & --                      & 1.3                          & 4                   &     1 & 1.0       & 377     &  301   &  229  \\
T36   & 1.7                     & 5.4                     & 1.3                          & 4                   &     2 & 0.1       &  5.8    &   4.5  &  3.3  \\
T37   & 1.7                     & 5.4                     & 1.3                          & 4                   &     2 & 0.2       & 24      &  18    &   15  \\
\enddata
\tablenotetext{a}{Identifier of the run (for each run, 10000 BSSs have been simulated).}
\tablenotetext{b}{$r_{\rm min,\,{}MT}$ and $r_{\rm max,\,{}MT}$ are the minimum and the maximum three-dimensional radius within which MT-BSSs are generated in the simulations. The minimum and the maximum three-dimensional radius within which COL-BSSs are generated in the simulations are not listed in the Table, because they are the same in all the simulations (i.e., $r_{\rm min,\,{}COL}=0$ and $r_{\rm max,\,{}COL}=r_{\rm c}$, respectively).}
\tablenotetext{c}{Mass of a simulated  BSS. In the simulations, $m_{\rm BSS}$ is the same for MT-BSSs and COL-BSSs.}
\tablenotetext{d}{Lifetime of a simulated  BSS. In the simulations, $t_{\rm life}$ is the same for MT-BSSs and COL-BSSs.}
\tablenotetext{e}{Kick velocity for COL-BSSs.} 
\tablenotetext{f}{Fraction of simulated COL-BSSs with respect to the total population of BSSs.} 
\tablenotetext{g}{$\chi{}^2_{\rm RGB}$, $\chi{}^2_{\rm SGB}$ and $\chi{}^2_{\rm HB}$ are the non-reduced $\chi{}^2$ of the simulated $N_{\rm BS\_RGB}$, $N_{\rm BS\_SGB}$ and $N_{\rm BS\_HB}$.}
\end{deluxetable*}


\section{Comparison with the Sculptor dSph}\label{sec:discu}

The results discussed in the previous sections provide evidence that the
two isolated dwarf galaxies Cetus and Tucana host a significant population of BSSs,
whose properties strongly suggest that they are the product of mass exchange
in binary systems. We did not find any evidence supporting the hypothesis that 
part of the objects populating the blue plume are intermediate-age or young MS stars.
On the basis of these findings, it appears that both Cetus and 
Tucana are similar to Ursa Minor, Draco \citep{mapelli07}, and Sculptor \citep{mapelli09}. 
In the following, we first focus on the comparison of Tucana and Sculptor, two 
dwarfs which show close similarities in terms of their SFH and present-day stellar content.


\begin{figure}
\resizebox{8truecm}{9truecm}{\includegraphics{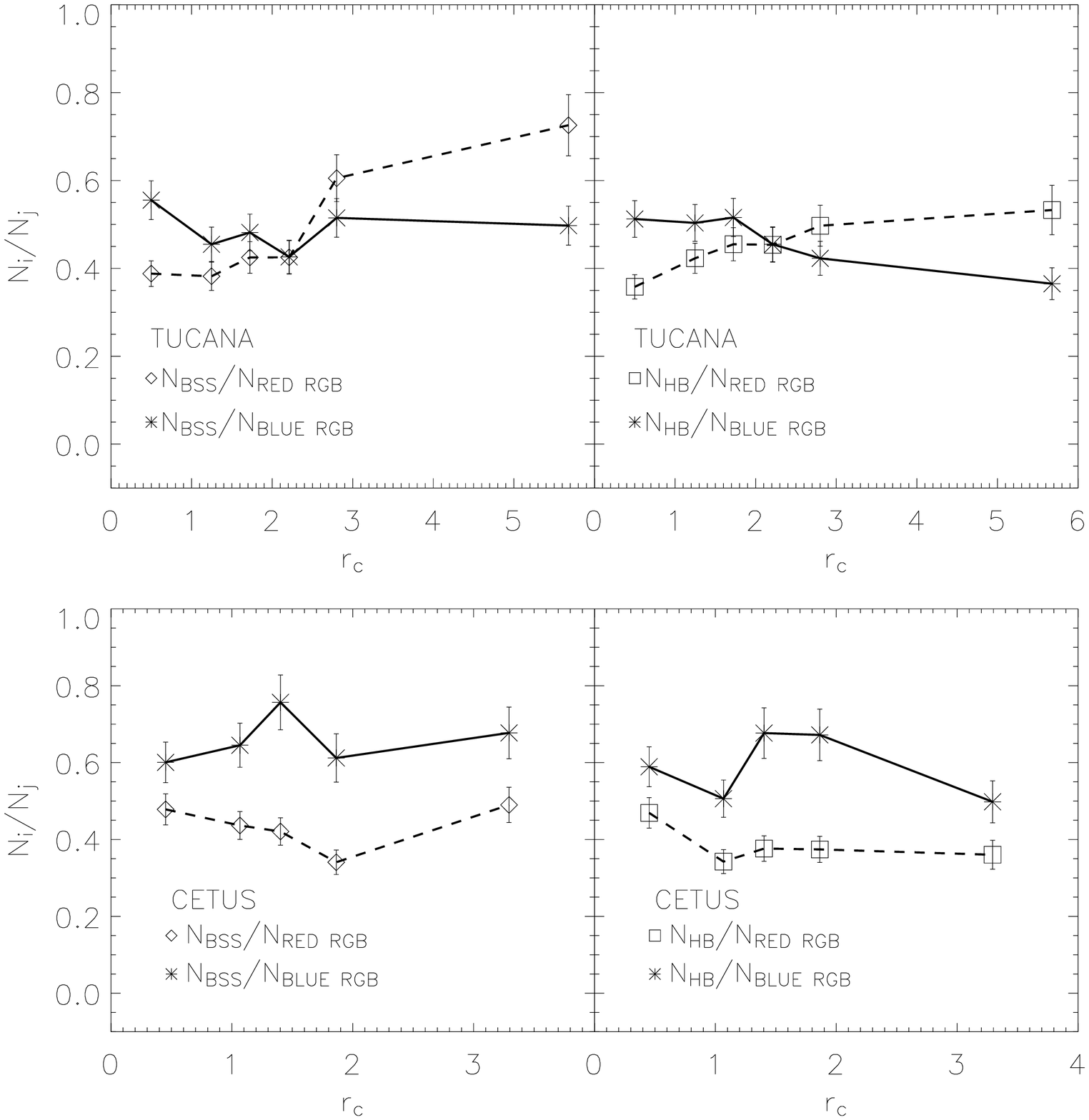}}
 \caption{Ratios as a function of galactocentric distance for different samples of stars.
 BSSs (left) and HB stars (right) are compared to red (assterisks, dashed line) and blue 
 (open symbols and solid line) RGB stars for Tucana (top) and Cetus (bottom). 
 }
 \label{fig:grad_rad2}
\end{figure}


Sculptor is a well established case of a dSph galaxy hosting two different old populations.
\citet{majewski99} first noted evidence of two RGB bumps in Sculptor. 
\citet{tolstoy04} presented a photometric and spectroscopic (235 RGB stars) investigation 
of this galaxy, showing the presence of strong radial gradients and two stellar 
populations with different kinematic and chemical properties.
Moreover, the radial distributions of the red and blue HB components are 
correlated with the properties of the RGB stars, suggesting that red stars are more
centrally concentrated, more metal-rich, and have a higher velocity dispersion, while the 
blue component has a broader distribution, lower metallicity, and a smaller velocity dispersion.

\citet{bernard08} showed the first evidence of stellar radial gradients in Tucana;
after dividing both the HB and the RGB into red and blue components, the number 
of stars belonging to the red components decreases faster than the corresponding 
blue sequence with increasing galaxy radius. Moreover, the analysis of the properties 
of RR Lyrae variables stars \citep{bernard09}, the SFH \citep{monelli10c}, and the
double RGB bump \citep{monelli10a} strongly support the coexistence of two old 
($>$ 10 Gyr) populations in Tucana, with slightly different ages, metallicities, 
and spatial distributions.  

\begin{figure*}
\resizebox{8truecm}{7truecm}{\includegraphics{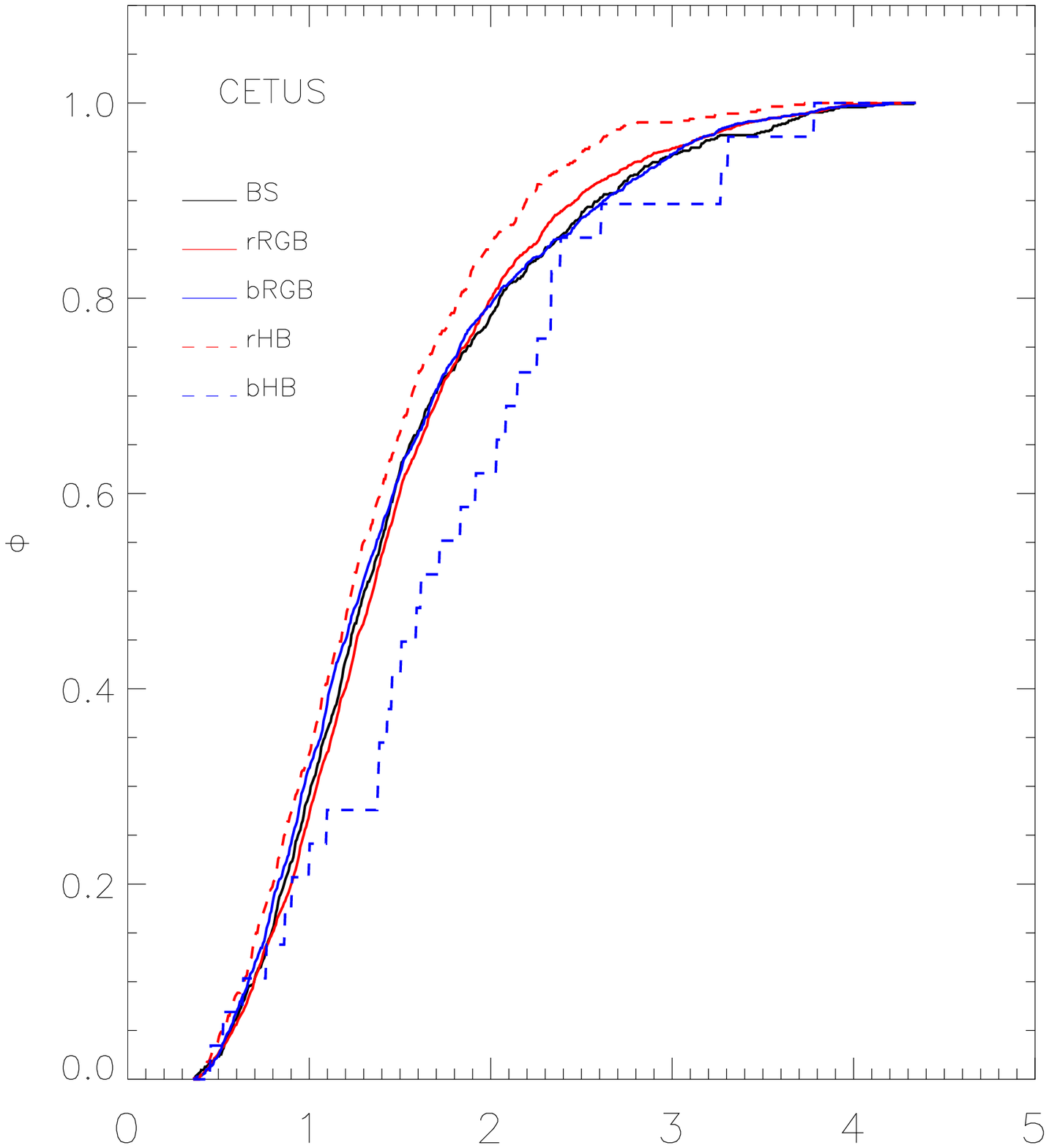}}
\resizebox{8truecm}{7truecm}{\includegraphics{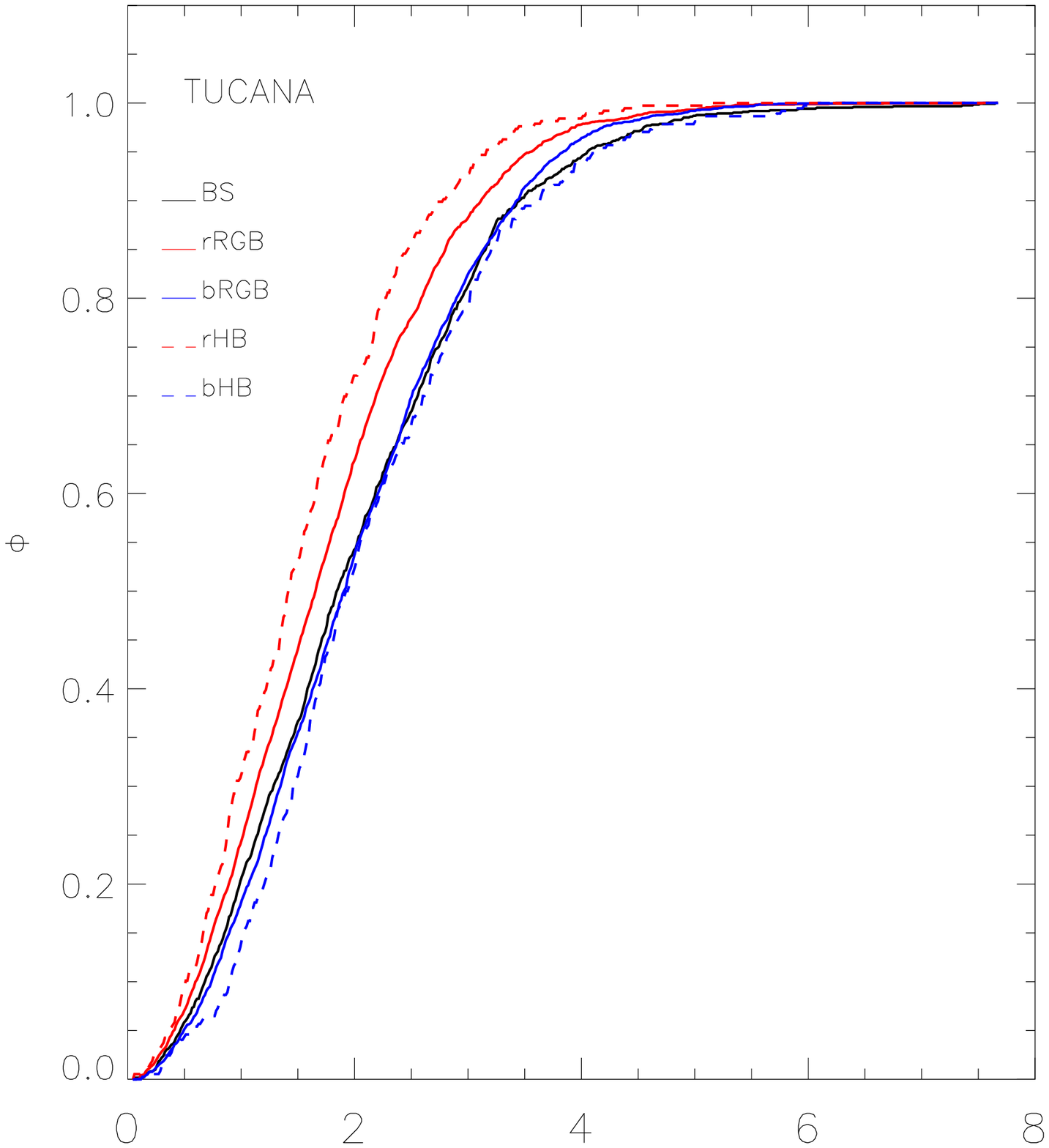}}
 \vskip 0.4cm
 \caption{Normalized cumulative distributions for the samples of BSSs (black solid line)
 compared to red and blue RGB stars (red and blue solid lines), and red and blue HB stars
 (red and blue dashed). In the Case of Cetus (left panel), the BSSs are more concentrated
 than the blue HB stars, and less concentrated than the red HB stars. In the case of 
 Tucana (right panel), the BSSs are less concentrated than both the red RGB and HB samples,
 while they are marginally more concentrated than the blue components.}
 \label{fig:cumul2}
\end{figure*}


Figure \ref{fig:scltuc} compares the ratio of the number of BSSs versus red (asterisks)
and blue (squares) HB stars as a function of galactocentric radius for the two galaxies.
The values for Sculptor are taken from Table 2 of \citet{mapelli09}. Interestingly, 
in both cases the BSSs are more centrally concentrated than the blue component 
of the HB, and less concentrated than the red component. We note also that, for each galaxy, 
the two curves cross at a similar galactocentric distance, $\approx 1-1.6$ $r_c$. 
In the case of Tucana, this trend is confirmed also for the BSSs and the HB stars 
when compared to red and blue RGB stars (Figure \ref{fig:grad_rad2}, top left and top right, 
respectively). Therefore, if we compare the global N$_{\rm BSS}$/N$_{\rm HB}$ ratio
in Figure \ref{fig:grad_rad} with Figure \ref{fig:scltuc} and Figure \ref{fig:grad_rad2}
(top panels), it can be inferred that the radial trends of the 
BSSs are indeed driven by the fast decrease of the red HB stars in the central regions, 
associated with the younger, more metal-rich component, while the increase at larger radii
is driven by the more homogeneously distributed bluest stars.  
This means that the peculiar radial trend of the candidate BSSs when they are normalized 
to the HB stars is not due to a peculiar distribution of these objects inside the galaxy 
but only to the existence of a significant metallicity gradient in Tucana, a gradient that 
is well traced by both the red and blue HB stellar populations.

The similarities between these two galaxies are particularly intriguing because of
the remarkably different environmental conditions: Tucana spent most of its life in
isolation, while Sculptor is a close satellite of the Milky Way. If all the BSSs
are descendants of primordial binaries, this suggests that the different environments
of the two galaxies did not alter their binary distributions.

In contrast, Figure \ref{fig:grad_rad2} (bottom panels) shows that the properties of 
different stellar samples in Cetus appear more homogeneous. This would be in agreement with 
other findings, as the properties of the RR Lyrae variable stars and of the RGB bump, which
indicate a single dominant population characterized by a more uniform distribution
of ages and metallicities. 

Interesting features appear when comparing the distribution of the BSSs with those of 
the red and blue stars of both the RGB and HB. We defined these samples as follows. The
region selecting the RGB stars was simply split in two, similarly to what has been done 
for the analysis of the RGB bump in \citet{monelli10a}. The red and blue HB are defined
considering only the stars redder than $m_{F475W}-m_{F814W} > 0.99$ and bluer than 
$m_{F475W}-m_{F814W} < 0.46$ in the HB region, therefore mostly avoiding the RR Lyrae
instability strip. Figure \ref{fig:cumul2} compares the distributions of these four
samples with the BSSs for both Tucana and Cetus. In the case of Cetus, we find that the BSSs
are significantly less concentrated than the red HB and more concentrated than the blue 
HB, while they more closely follow the distributions of both of the RGB components. In the 
case of Tucana, the BSSs closely follow the distributions of both blue components
(possibly more concentrated in the central regions), while they are significantly 
less concentrated than the red sequences. 

We have demonstrated that the populations of BSS candidates in Tucana and Cetus are
consistent with all predictions of creation from MT-BSSs. In fact, it would be 
unprecedented to have a truncation in star formation without an associated population
of MT-BSSs.  Unfortunately, our limited knowledge of the evolution of binary 
stellar systems prevents us from making accurate predictions of the expected numbers
of MT-BSSs.  In sum, we expect to see MT-BSSs in dSph galaxies, and we detect them
with consistent properties.  The remaining question is whether some fraction of the
BSS candidates could still be due to newly formed MS stars.  The inconsistency
with the age-metallicity relationship could be due to infall of a gas cloud with
a low metallicity coincident with that of the main population which then
produces subsequent star formation.  However, the truncation at 
luminosities corresponding to twice the TO mass implies a second unlikely coincidence
if some of the BSS candidates were true MS stars.  While very difficult to rule
out such a scenario completely, it is clear that newly formed MS stars are not
necessary to explain all of the observations.

\section{Summary and conclusion}\label{sec:conclu}

In this paper, we have presented the analysis of candidate BSSs in two isolated 
dSphs in the periphery of the Local Group, Cetus and Tucana. Deep HST/ACS data allowed 
us to identify 940 and 1214 BSS candidates in Cetus and Tucana, respectively.
Different indicators suggest that these are true BSSs; and, thus, there are no young main
sequence stars in either galaxy. The first evidence comes from the 
SFH we derived for both objects \citep{monelli10b, monelli10c}.  Not allowing for
the presence of BSSs, then a small population
of relatively young (3 to 5 Gyr old) stars is detected, with metallicities 
similar to the oldest most metal-poor stars, and thus, much
lower than expected from the age-metallicity relation of the dominant population.
The comparison with stellar tracks
presented in this paper suggests that the BSSs have typical masses $> 1{}M_{\odot}$, and 
smaller than $\sim1.5{}M_{\odot}$, in agreement with the fact that the maximum 
expected mass of a BSS is twice the mass presently evolving at the TO ($\sim 0.8\,{}M_\odot{}$). 
Moreover,
the limited spread in metallicity found for the BSS populations, perfectly in agreement 
with that of turn-off and subgiant stars, suggests that the use of stellar evolution 
models for single stars does not introduce strong biases in the analysis.

The analysis of the luminosity functions and radial distributions support the
BSS interpretation.  On one hand,
we showed that the luminosity function does not change as a function of radius,
and in particular the brightest (youngest) stars do not appear more centrally
concentrated, as one would expect from a young main sequence. On the other hand,
the radial profiles do not show any central peak, as observed in globular
clusters due to COL-BSS stars or as it would be in the case of a young component.
 
Thus, the present analysis also suggests that the BSSs in Cetus and Tucana formed
from primordial binaries and, as in the case of other dSphs, COL-BSSs are unlikely
to form. This conclusion is strongly supported by the dynamical 
simulations we performed, taking into account the properties of the 
dwarfs under scrutiny. These simulations indicate that the observed BSS distributions
are well reproduced when only MT-BSSs are taken into account, while the fit worsens
when even a small number of COL-BSSs ($\sim10$\%) is included in the dynamical 
simulation.

We can therefore safely conclude that it is highly likely that Cetus and Tucana host 
a genuine population of BSS stars, and we confirm the conclusions presented in previous 
studies that both galaxies did not experience any star formation episodes in the last 
8 Gyr.  It is particularly interesting that the positions of Cetus and Tucana in the 
F$_{BS}$ vs.\ $M_V$ plane fit very well in the relation presented by \citet{momany07}.
This, in turn, could be a powerful instrument to identify young main sequence stars
in resolved spheroidal galaxies, discriminating from systems with only old populations
from those also hosting younger stars.

\acknowledgments
The authors thank the anonymous referee for the useful comments, Yazan Momany for 
sharing the data used in Figure~\ref{fig:fm}, and
Giuliana Fiorentino for useful discussions on the ACs.
Support for this work was provided by NASA through grant GO-10515
from the Space Telescope Science Institute, which is operated by
AURA, Inc., under NASA contract NAS5-26555, the IAC (grant 310394), the
Education and Science Ministry of Spain (grants AYA2004-06343, AYA2007-3E3507,
and AYA2010-16717). S.C. acknowledges the financial support of INAF
through the PRIN INAF 2009 (P.I.: R. Gratton).
This research has made use of NASA's Astrophysics Data System
Bibliographic Services and the NASA/IPAC Extragalactic Database
(NED), which is operated by the Jet Propulsion Laboratory, California
Institute of Technology, under contract with the National Aeronautics
and Space Administration.

{\it Facilities:} \facility{HST (ACS)}.

\end{document}